\documentclass[trackchanges,twocolumn,resetfootnote]{aastex701}
\usepackage{booktabs}
\usepackage{fancyhdr}
\usepackage{threeparttable}
\usepackage{siunitx}
\usepackage{amsmath,amsfonts,amssymb}
\usepackage{mathtools}
\usepackage{physics}
\usepackage{bm}
\usepackage{enumerate}
\usepackage{makecell}

\defcitealias{maltsevExplodability2025}{M25}
\defcitealias{ebinger2019PUSHing}{E19}
\defcitealias{ebingerPUSHing2020}{E20}

\begin{document}

\title{Constraints on the Metallicity-dependent Explodability of Massive Stars from Galactic Chemical Evolution: Toward Alleviating the Red Supergiant Problem}

\author[orcid=0009-0007-8748-2717,gname=Sojun,sname='Ono']{Sojun Ono}
\affiliation{Department of Astronomy, Kyoto University, Kitashirakawa-Oiwake-cho, Sakyo-ku, Kyoto 606-8502, Japan}
\email[show]{ono@kusastro.kyoto-u.ac.jp}  

\author[orcid=0000-0003-2611-7269,gname=Keiichi,sname='Maeda']{Keiichi Maeda} 
\affiliation{Department of Astronomy, Kyoto University, Kitashirakawa-Oiwake-cho, Sakyo-ku, Kyoto 606-8502, Japan}
\email{keiichi.maeda@kusastro.kyoto-u.ac.jp}

\author[orcid=0000-0002-7043-6112,gname=Akihiro,sname='Suzuki']{Akihiro Suzuki} 
\affiliation{Research Center for the Early Universe, The University of Tokyo, 7-3-1 Hongo, Bunkyo-ku, Tokyo 113-0033, Japan}
\email{akihiro.suzuki@resceu.s.u-tokyo.ac.jp}

\begin{abstract}
The explodability of massive stars, namely whether they undergo core-collapse supernovae (CCSNe) or form black holes (BHs), strongly influences galactic chemical evolution (GCE). Details of the explodability are still controversial, but realistic predictions including metallicity-dependence are becoming available through stellar-evolution and explosion calculations. In the present work, we implement recently-proposed metallicity-dependent explodability prescriptions into a GCE framework. We show that the physics-motivated explodability prescriptions reproduce the key observed abundance trends. Further, within uncertainties of the explodability models, the GCE model provides important constraints on the region of the BH formation in the mass-metallicity space. Guided by these findings, we further construct a simplified form of the metallicity-dependent explodability designed to alleviate the red supergiant (RSG) problem and explore its compatibility with GCE constraints. We find that such a solution exists, if (1) the net outflows from the system are negligible/absent, and (2) the transition of the explodability takes place at sub-solar metallicity. These results demonstrate that GCE can provide meaningful constraints on massive-star explodability and that explodability prescriptions capable of addressing the RSG problem can be constructed without violating chemical-evolution observables. We also show that a metallicity-dependent initial mass function can improve agreement with observations; this effect becomes important once coupled with the metallicity-dependent explodability.
\end{abstract}



\section{Introduction}\label{sec:introduction}

Heavy elements in the Universe were not produced during the Big Bang; rather, they are synthesized in stars and subsequently dispersed into the interstellar medium (ISM) through stellar winds and supernova (SN) explosions. Stellar feedback, in particular the mass loss and SN ejecta, is essential for understanding the chemical enrichment history of galaxies such as the Milky Way \citep[e.g.,][]{timmes1995Galactic, pagel1997Nucleosynthesis, matteucci2001Chemical, kobayashiOrigin2020}, commonly referred to as Galactic Chemical Evolution (GCE). Conversely, one can impose constraints on the underlying stellar-evolution and supernova-physics assumptions through GCE models.

Core-collapse SNe (CCSNe) represent one of the most prominent endpoints of massive-star evolution, ejecting newly synthesized metals into the ISM. However, not all massive stars result in successful CCSNe: a non-negligible fraction of them are thought to undergo failed explosions and form black holes (BHs), thereby contributing little or no metal enrichment. The stellar property that governs whether the collapse results in a successful or failed explosion is commonly termed explodability, and it constitutes one of key uncertainties in GCE models.

However, in many GCE studies, little attempt has been made to constrain explodability \citep[e.g.,][]{suzukiConstraining2018,vincenzo2018Extragalactic,tsujimotoGalactic2022}.  From the perspective of stellar evolution and explosion modeling, various numerical studies have attempted to predict explodability across progenitor properties \citep[e.g.,][]{sukhbold2016CORECOLLAPSE}, recently expended to explore how explodability may depend on metallicity \citep[e.g.,][]{ebingerPUSHing2020, maltsevExplodability2025}. These one-dimensional models have been calibrated against computationally-expensive multidimensional neutrino-driven SN explosion simulations \citep[e.g.,][]{melson2015NEUTRINODRIVEN, lentz2015THREEDIMENSIONAL, burrows2020Overarching}, and cover a wide range of masses and metallicities that can be applicable in GCE calculations as we explore in the present study. \footnote{\citet{jost2025Neutrinodriven} calculated neutrino-driven CCSN yields with three metallicities and applied them to GCE calculations. However, explodability they adopted is such that non-explosions occur preferentially for low-compactness progenitors (often low-mass), whereas higher-compactness (typically higher-mass) progenitors explode more readily; this differs from the intermediate-mass ($\sim18$--$40\,M_\odot$) failed-SN window adopted in this work.}

An important observational tension related to the fate of massive stars is the so-called \emph{red supergiant (RSG) problem}. It refers to the apparent lack, in the local Universe, of identified Type~IIP SN RSG progenitors with zero-age main-sequence (ZAMS) masses above $\sim 18\,M_\odot$, despite stellar-evolution expectations that more massive RSGs should also undergo core collapse \citep{smarttProgenitors2009}. A number of explanations have been proposed; one possibility is that higher-mass progenitors preferentially undergo failed explosions and collapse to BHs \citep{kochanek2008Survey}.

This motivates investigation of whether the assumption that CCSNe occur only for $M_{\rm ZAMS}\lesssim 18\,M_\odot$ can remain consistent with GCE constraints. \citet{suzukiConstraining2018}, assuming that this is the case irrespective of the metallicity (i.e., metallicity-independent explodability prescription), pointed out that this leads to an oxygen abundance inconsistent with observations. A small description added in the above sentence. This stems from substantial contribution by stars with $M_{\rm ZAMS}\sim 20$--$40\,M_\odot$ to the net oxygen production; if these stars predominantly form BHs without ejecting oxygen-rich ejecta, the integrated oxygen enrichment is reduced (see Appendices~\ref{sec:comp_yields} and \ref{sec:yield_O_Fe}).

In addition to the metallicity-dependent explodability and the RSG problem, there are related topics. Type~Ia SNe (SNe~Ia) are likely composed of contributions from both Chandrasekhar-mass (Ch SNe~Ia) and sub-Chandrasekhar-mass (sub-Ch SNe~Ia) white dwarf explosions. These two channels exhibit distinct nucleosynthetic yield patterns. Consequences of such progenitor mixtures to GCE have been actively discussed in recent studies \citep{kobayashiNew2020, eitnerObservational2023}. Here, once the baseline CCSN yields would be changed, the arguments related to the SN Ia progenitors may also need to be re-investigated. In addition, there is growing interest in metallicity-dependent effects other than explodability. Recent work has begun to constrain the initial mass function (IMF) even in low-metallicity environments, including Population~II/III contexts \citep[e.g.,][]{chon2024Impact}, and it is often argued that the IMF becomes more top-heavy at lower metallicity; an increased fraction of massive stars could modify enrichment histories. Moreover, explodability also affects the CCSN rate, which might be connected to the ``missing SN'' problem; the observed CCSN rate appears to be smaller (by a factor of $\sim 2$) than the rate expected from the cosmic star-formation rate (SFR) \citep{horiuchiCOSMIC2011}. Uncertainty in the mass range to produce CCSNe (i.e., explodability) is one factor that may be relevant to this discrepancy.

In this paper, we investigate possible GCE constraints on the explodability, by computing GCE models with metallicity-dependent explodability motivated by recent stellar-evolution calculations. In Section~\ref{sec:methods}, we describe the setup of our GCE models and introduce the metallicity-dependent explodability models proposed from SN explosion calculations. After that, we presents the GCE results, focusing on the oxygen evolution. Based on these results, Section~\ref{sec:simplified_model} presents potential ways in which the RSG problem could be alleviated; for this purpose, we introduce a simplified prescription for explodability. In Section~\ref{sec:discussion}, we discuss several related topics. Finally, in Section~\ref{sec:conclusion}, we summarize our main results and conclusions and outline future prospects.

Throughout this paper, we convert redshift $z$ to cosmic time $t$ assuming a $\Lambda$CDM cosmology. We adopt the $z$--$t$ relation implemented by \citet{wright2006Cosmology} and use the cosmological parameters inferred from \citet{aghanim2020Planck}; the Hubble constant $H_0 = 67.66\,\mathrm{km\,s^{-1}\,Mpc^{-1}}$, the dark-energy density parameter $\Omega_\Lambda = 0.6889$, and the matter density parameter $\Omega_\mathrm{m} = 0.3111$.

\begin{deluxetable}{lccccccc}
\tablewidth{0pt}
\tablecaption{The settings of each GCE model \label{table:GCEparams}}
\tablehead{
    \colhead{Model} &
    \colhead{Explodability} &
    \colhead{$\epsilon_\mathrm{out}$} &
    \colhead{$\tau_\mathrm{in}$ [Gyr]} &
    \colhead{$\tau_\mathrm{s}$ [Gyr]} &
    \colhead{$\dot{M}_\mathrm{in,0}$ [$M_\odot\,\mathrm{yr}^{-1}$]} &
    \colhead{IMF} &
    \colhead{Ch/sub-Ch}
}
\startdata
\cutinhead{Base models}
Base-100    & CCSN: $8\text{--}100\,M_\odot$ & $0.75$ & $5.0$ & $3.0$ & $20$ & Kroupa & 100/0 \\
Base-40     & CCSN: $8\text{--}40\,M_\odot$  & $0.75$ & $5.0$ & $3.0$ & $20$ & Kroupa & 100/0 \\
Base-18     & CCSN: $8\text{--}18\,M_\odot$  & $0.75$ & $5.0$ & $3.0$ & $20$ & Kroupa & 100/0 \\
\cutinhead{\makecell[c]{group Ex (see Section~\ref{sec:methods})}}
Ex-M25-S    & ``Single'' in \citetalias{maltsevExplodability2025} & $0.75$ & $5.0$ & $3.0$ & $20$ & Kroupa & 100/0 \\
Ex-M25-B    & ``Case B'' in \citetalias{maltsevExplodability2025} & $0.75$ & $5.0$ & $3.0$ & $20$ & Kroupa & 100/0 \\
Ex-PUSH     & \citetalias{ebinger2019PUSHing} and \citetalias{ebingerPUSHing2020} & $0.75$ & $5.0$ & $3.0$ & $20$ & Kroupa & 100/0 \\
\cutinhead{\makecell[c]{group Ex-BH (see Section~\ref{sec:methods})}}
Ex-M25-S-BH & \makecell[c]{``Single'' in \citetalias{maltsevExplodability2025} \\ w/o mixed regions} & $0.75$ & $5.0$ & $3.0$ & $20$ & Kroupa & 100/0 \\
Ex-M25-B-BH & \makecell[c]{``Case B'' in \citetalias{maltsevExplodability2025} \\ w/o mixed regions} & $0.75$ & $5.0$ & $3.0$ & $20$ & Kroupa & 100/0 \\
\cutinhead{\makecell[c]{Simplified models (see Section~\ref{sec:simplified_model})}}
Reference model ($\epsilon_\mathrm{out}=0.75$) & Eq.~\eqref{eq:Mth} & $0.75$ & $5.0$ & $3.0$ & $20$ & Kroupa & 100/0 \\
No outflow model ($\epsilon_\mathrm{out}=0.00$) & Eq.~\eqref{eq:Mth} & $0.00$ & $5.0$ & $3.0$ & $8$  & Kroupa & 100/0 \\
\cutinhead{\makecell[c]{+ sub-Ch SN Ia (see Section~\ref{section:sub-CH})}}
Ex-M25-S + sub-Ch SN Ia      & ``Single'' in \citetalias{maltsevExplodability2025} & $0.75$ & $5.0$ & $3.0$ & $20$ & Kroupa & 50/50 \\
\cutinhead{\makecell[c]{+ Chon IMF (see Section~\ref{section:MD-IMF})}}
Base-100 + Chon IMF    & CCSN: $8\text{--}100\,M_\odot$ & $2.00$ & $3.5$ & $2.5$ & $71$ & Chon & 100/0 \\
Base-18 + Chon IMF    & CCSN: $8\text{--}18\,M_\odot$ & $2.00$ & $3.5$ & $2.5$ & $71$ & Chon & 100/0 \\
Ex-M25-S + Chon IMF    & ``Single'' in \citetalias{maltsevExplodability2025} & $2.00$ & $3.5$ & $2.5$ & $71$ & Chon & 100/0 \\
Ex-M25-B + Chon IMF    & ``Case B'' in \citetalias{maltsevExplodability2025} & $2.00$ & $3.5$ & $2.5$ & $71$ & Chon & 100/0 \\
Ex-PUSH + Chon IMF    & \citetalias{ebinger2019PUSHing} and \citetalias{ebingerPUSHing2020} & $2.00$ & $3.5$ & $2.5$ & $71$ & Chon & 100/0 \\
\enddata
\tablecomments{
\citetalias{maltsevExplodability2025}, \citetalias{ebinger2019PUSHing}, and \citetalias{ebingerPUSHing2020} refer to \citet{maltsevExplodability2025}, \citet{ebinger2019PUSHing}, and \citet{ebingerPUSHing2020}, respectively.
}
\end{deluxetable}


\section{GCE Models Adopting the Metallicity-dependent Explodability}\label{sec:methods}


\subsection{Galactic Chemical Evolution}\label{section:main-method}

We adopt the GCE code of \citet{suzukiConstraining2018} as the baseline of this study. Although multizone GCE models are widely used and can include the effects of stellar migrators \citep[e.g.,][]{tsujimotoGalactic2022}, we adopt a one-zone framework to isolate the impact of metallicity-dependent explodability. We consider 31 chemical species, spanning from H to Ga. For the IMF, we use the Kroupa IMF \citep{kroupaVariation2001}\footnote{A set of additional GCE calculations with a metallicity-dependent IMF is presented in Section~\ref{section:MD-IMF}.}. The lower and upper limits of the stellar mass range are set to $M_l=0.08\,M_\odot$ and $M_u=100\,M_\odot$, respectively. Unless otherwise stated, we use the same parameter settings as in \citet{suzukiConstraining2018}.

We denote the gas mass in the galaxy by $M_{\mathrm{g}}$. The initial gas mass is set to $M_{\mathrm{g},0}=10^{7}\,M_\odot$ at $t=0\,\mathrm{Gyr}$. The mass fraction of element $i$ is denoted by $X_i$. With the hydrogen and helium mass fractions further defined as $X \equiv X_{\mathrm{H}}$ and $Y \equiv X_{\mathrm{He}}$, the gas metallicity is given as $Z = 1 - X - Y$. At $t=0\,\mathrm{Gyr}$, we assume a primordial composition with $X=0.75$, $Y=0.25$, and $Z=0.00$. For the solar metallicity, we consistently adopt $Z_\odot = 0.0142$ \citep{asplundChemical2009} throughout this paper. Some of the referenced yield sets and models assumed different values of $Z_\odot$; nevertheless, we report metallicities using this common convention. This choice does not affect the numerical implementation, because the calculation is performed using metallicity expressed as the absolute mass fraction $Z$, not $Z/Z_\odot$. We also adopt the solar abundance of individual elements from \citet{asplundChemical2009}.

The evolution of the gas content within the system, $\dot M_{\mathrm{g}}$, is controlled by the three processes; inflow, star formation, and outflow. We assume that the gas inflow rate follows an exponential decay in time. Specifically, we adopt an inflow timescale of $\tau_{\mathrm{in}}=5\,\mathrm{Gyr}$. With the initial inflow rate of $\dot{M}_{\mathrm{in},0}=1.0\times10^{11}\,M_\odot/\tau_{\mathrm{in}}=20\,M_\odot\,\mathrm{yr}^{-1}$ at $t = 0$ Gyr, the evolution of the inflow rate is given by
\begin{align}
    \dot{M}_{\mathrm{in}} = \dot{M}_{\mathrm{in},0}\,e^{-t/\tau_{\mathrm{in}}} .
\end{align}
We assume that the inflowing gas maintains the primordial composition ($X=0.75$, $Y=0.25$, and $Z=0.00$). The SFR is modeled with a star-formation timescale, $\tau_{\mathrm{s}}$; 
\begin{align}
    \mathrm{SFR} = \frac{M_{\mathrm{g}}}{\tau_{\mathrm{s}}} ,
\end{align}
where we adopt $\tau_{\mathrm{s}}=3\,\mathrm{Gyr}$. The outflow mass-loss rate is assumed to be proportional to the SFR \citep[e.g.,][]{matteucci2012Chemical, matteucci2021Modelling}:
\begin{align}
    \dot{M}_{\mathrm{out}} = \epsilon_{\mathrm{out}}\,\mathrm{SFR} ,
\end{align}
where we adopt $\epsilon_{\mathrm{out}}=0.75$. This parameter differs significantly from that adopted by \citet{suzukiConstraining2018} ($\epsilon_\mathrm{out}=2.5$), because in the present work, we calibrated the GCE parameters including $\epsilon_\mathrm{out}$ to satisfy several observational constraints (such as the metallicity distribution function, MDF, and the iron abundance at the epoch of the Sun's birth, $t=t_\odot=9.2\,\mathrm{Gyr}$.) for the introduced metallicity-dependent explodability. 

For the stellar yields except for the CCSN yields (which are discussed in Section \ref{section:Explodability-Methods}), we adopt the following prescriptions. For asymptotic giant branch (AGB) stars, we adopt yields from \citet{karakas2010Updated} for $M_{\mathrm{ZAMS}}\lesssim6.5\,M_\odot$. For super-AGB stars with $M_{\mathrm{ZAMS}}\gtrsim6.5\,M_\odot$, we use the yield tables of \citet{doherty2014Super,doherty2014Supera}. For SNe~Ia, we use the W7 model \citep{nomotoEvolution1984} as in \citet{suzukiConstraining2018}, adopting the nucleosynthetic yields from \citet{iwamotoNucleosynthesis1999}. For the delay-time distribution (DTD), we adopt observationally motivated forms \citep{graurTYPEIa2014, maozObservational2014}. Although \citet{chruslinska2024Trading} showed that variations in the minimum delay time can change the relation between the specific SFR and $[\mathrm{O/Fe}]$, we consider only a minimum delay time of $50\,\mathrm{Myr}$ in this work. All remaining SNe~Ia parameters follow \citet{suzukiConstraining2018}.
    
\begin{figure*}[htb!]
    \includegraphics[width=18cm]{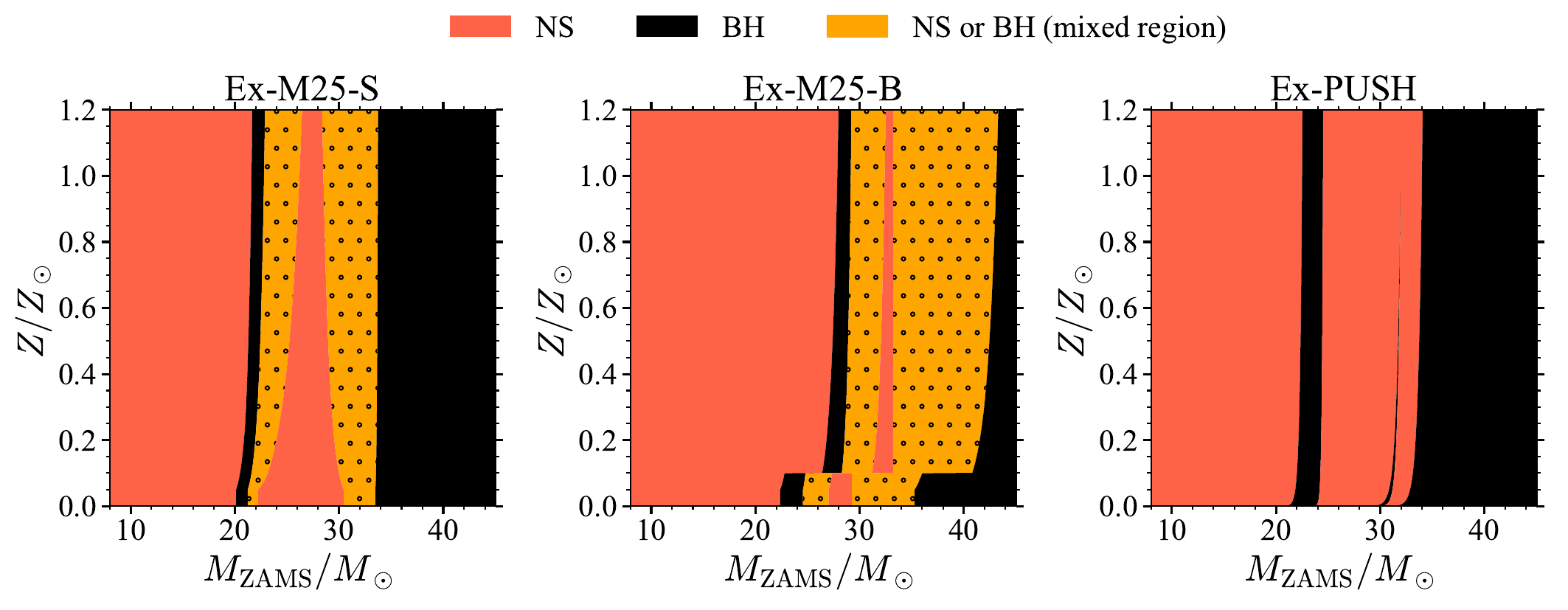}
    \caption{
    Schematic representation of the metallicity-dependent explodability adopted in group Ex. The left, center, and right panels show the explodability as functions of ZAMS mass and metallicity for models Ex-M25-S, Ex-M25-B, and Ex-PUSH, respectively. Red regions indicate successful CCSNe leaving neutron stars (NSs), while black regions denote direct black hole (BH) formation. The orange region with dotted patterns represents the mixed region where a NS-forming successful explosion takes place with a probability of 85\% while the remaining 15\% leads to a BH formation without ejecting metal-rich ejecta. 
    }
    \label{figure:explodability_schematic}
\end{figure*}


\subsection{CCSN Explodability and Yields}\label{section:Explodability-Methods}

In order to incorporate CCSN explodability into GCE, one should, in principle, adopt a fully self-consistent treatment in which explodability and nucleosynthetic yields are derived from the same SN explosion model. However, the oxygen abundance---the primary parameter investigated in this study---is relatively insensitive to differences among commonly used CCSN yield sets \citep[see][and Section~\ref{section:alpha_elements}]{suzukiConstraining2018}, because the oxygen yield is largely set by the pre-supernova stellar evolution (see Appendix~\ref{sec:comp_yields} in detail). Therefore, we approximate explodability and yields as separable ingredients and combine them even though they are not obtained by the same self-consistent calculation. This approximation should be used with caution, but it provides a practical way to explore the GCE impact of different explodability prescriptions.

In the present work, we adopt CCSN yield tables from \citet{chieffi2004Explosive}. They evolved stellar models from the pre-main sequence to core collapse and then computed the subsequent explosive nucleosynthesis using parametrically induced one-dimensional explosions. They provide CCSN yields for $Z=0,\,10^{-6},\,10^{-4},\,10^{-3},\,6\times10^{-3},\,\mathrm{and}\ 2\times10^{-2}$ with the progenitor ZAMS masses $M_\mathrm{ZAMS}=13.0,\,15.0,\,20.0,\,25.0,\,30.0,\,\mathrm{and}\ 35.0\,M_\odot$. Moreover, the CCSN yields are given as a function of mass coordinate for each model, allowing the ejected yields to be evaluated for different choices of the ejected $^{56}\mathrm{Ni}$ mass. In GCE calculations, the ejected mass of $^{56}\mathrm{Ni}$ is a key parameter translated to the Fe yield. This is determined by the so-called `mass cut' that separates the material ejected into the ISM and that collapsed onto a central remnant. Because the mass cut can be varied in these yield tables, they can in principle be combined with any explosion model that predicts the remnant mass and explodability. Although this flexibility requires caution, it allows the GCE constraints derived in this work to be applied to different explosion prescriptions.

We adjust the mass cut, and hence the ejected $^{56}\mathrm{Ni}$ mass, so that it roughly follows the predictions of one-dimensional neutrino-based explosion models that are calibrated or parameterized to produce explosions \citep[e.g.,][]{sukhbold2016CORECOLLAPSE,ebinger2019PUSHing,ebingerPUSHing2020}. Specifically, we adopt a piecewise prescription for the ejected $^{56}\mathrm{Ni}$ mass: we set $M_{^{56}\mathrm{Ni}}=0.02\,M_\odot$ for $8\,M_\odot \leq M_\mathrm{ZAMS}\leq 14\,M_\odot$, and $M_{^{56}\mathrm{Ni}}=0.07\,M_\odot$ for $M_\mathrm{ZAMS}>14\,M_\odot$. The average $^{56}$Ni production adopted here is also consistent with the estimate through SN observations \citep[e.g.,][]{martinez2022Type}. We note, however, that the relationship between explodability and mass-cut remains uncertain, since self-consistent multi-dimensional CCSN simulations are not yet available over a sufficiently wide range of progenitor masses and metallicities. Our treatment should therefore be regarded as a simple, observationally motivated approximation rather than a definitive prediction of the mass cut.

We construct several models for the metallicity-dependent explodability based on the results of \citet{maltsevExplodability2025, muller2016Simple, mandel2020Simple} and \citet{ebinger2019PUSHing,ebingerPUSHing2020}. These include models Ex-M25-S, Ex-M25-B, and Ex-PUSH (collectively called \emph{group Ex}; Figure~\ref{figure:explodability_schematic}), and models Ex-M25-S-BH and Ex-M25-B-BH (\emph{group Ex-BH}). These models are described in the following subsections and Table~\ref{table:GCEparams}. All models adopt a fixed lower ZAMS-mass threshold of $8\,M_\odot$ for CCSNe \footnote{We have also performed the calculations by changing the lower ZAMS-mass threshold to $9\,M_\odot$ and $10\,M_\odot$ for CCSNe, and confirmed that our main conclusions would not be affected.}.


\subsubsection{Base models}\label{section:Base-models}

As a baseline reference, we define two GCE models, Base-100 and Base-18, in which the CCSN-producing mass ranges are assumed to be $8$--$100\,M_\odot$ and $8$--$18\,M_\odot$, respectively, irrespective of the metallicity.  These serve as reference cases corresponding to ``metallicity-independent explodability'', similar to the models discussed in \citet{suzukiConstraining2018}; Base-100 represents a canonical model widely used in the GCE study, while Base-18 represents a model in which the explodability (across the metallicity) is set to avoid the RSG problem in the local Universe. These models are used as comparison benchmarks for models with metallicity-dependent explodability.

\subsubsection{Models Ex-M25-S and Ex-M25-B}\label{section:M25_explodability}

\citet{maltsevExplodability2025} proposed a metallicity-dependent explodability based on the CO-core mass $M_\mathrm{CO}$. After computing the explosion/non-explosion boundaries in CO-core mass ($M_\mathrm{CO,bound}$) at $Z=Z_\odot/10$ and $Z_\odot$, they extrapolated $M_\mathrm{CO,bound}$ to other metallicities by imposing a logarithmic dependence, $M_\mathrm{CO,bound}\propto \log(Z/Z_\odot)$. However, to avoid an unphysical divergence of $M_\mathrm{CO,bound}$ at $Z\rightarrow 0$, we provisionally fix $M_\mathrm{CO,bound}$ to its value at $Z=Z_\odot/20$ for all $Z<Z_\odot/20$. We note that \citet{maltsevExplodability2025} argued that this explodability scheme might solve the RSG problem; however, whether it is simultaneously consistent with observed GCE is a separate question, and we explore this topic in the present work. 

In this work, we use the results labeled ``Single'' and ``Case B'' in \citet{maltsevExplodability2025}. The ``Single'' represents explodability for single-star evolution. The ``Case B'' represents a binary evolution in which the hydrogen-rich envelope is removed between the end of the main sequence and core helium exhaustion. We denote the GCE models using these two explodability prescriptions as Ex-M25-S (Single) and Ex-M25-B (Case B). Both of these models belong to group Ex.

The left and middle panels of Figure~\ref{figure:explodability_schematic} show the explodability in models Ex-M25-S and Ex-M25-B, respectively. The explodability in these models has mixed regions in which a neutron star (NS)-forming successful explosion takes place with a probability of 85\% while the remaining 15\% leads to a fallback BH formation. In the fallback BH formation, a large fraction of the core materials fallback to a BH, and thus no newly-formed metal is ejected into ISM in our calculations. We discuss yields from BH-formation events further in Section~\ref{section:BH_yields}.

Although explodability is more naturally characterized in terms of $M_\mathrm{CO}$ rather than $M_\mathrm{ZAMS}$ \citep{maltsevExplodability2025}, our GCE calculations and CCSN yields are formulated as functions of $M_\mathrm{ZAMS}$. We therefore convert the explodability expressed in terms of $M_\mathrm{CO}$ into an equivalent dependence on $M_\mathrm{ZAMS}$. For this conversion, we adopt the relationship between $M_\mathrm{CO}$ and $M_\mathrm{ZAMS}$ presented in \citet{schneider2023Bimodal}, as also introduced in \citet{maltsevExplodability2025}. Specifically, for models based on Case B in \citet{maltsevExplodability2025}, we use the corresponding Case B relation from \citet{schneider2023Bimodal}; for the other models, we use their Single-star relation. Because the mapping between $M_\mathrm{CO}$ and $M_\mathrm{ZAMS}$ is available only at $Z=Z_\odot$ and $Z=Z_\odot/10$, we apply the $Z=Z_\odot/10$ mapping for all $Z\leq Z_\odot/10$ and the $Z=Z_\odot$ mapping for $Z> Z_\odot/10$. This choice introduces an artificial discontinuity in the explodability of model Ex-M25-B at $Z=Z_\odot/10$ (Figure~\ref{figure:explodability_schematic}); however, as described later, we impose the same prescription when assigning the yields, so the discontinuity is internally consistent and therefore not problematic. We note that this remapping is approximate because the relation is sensitive to the individual binary mass-transfer history. A population-synthesis-based treatment is left for future work.

The explodability in the mixed region is less certain than in the other regions, and thus we also consider an empirical/extreme case in which the entire mixed regions form BHs, i.e., a case that maximizes the BH formation probability. We refer to these `without-mixed-region models' as Ex-M25-S-BH (Single) and Ex-M25-B-BH (Case B). These two models constitute group Ex-BH. Because group Ex-BH has only a very small CCSN-producing mass range above $18\,M_\odot$ at $Z\approx Z_\odot$, it may provide hints toward resolving the RSG problem.

When using the models Ex-M25-B or Ex-M25-B-BH, additional care is required in assigning yields. Because nucleosynthesis of metals occurs primarily in the stellar core, the yields are expected to depend more strongly on $M_\mathrm{CO}$ than on $M_\mathrm{ZAMS}$. Accordingly, for Case B we approximate binary-star yields by mapping $M_\mathrm{ZAMS}$ to $M_\mathrm{CO}$, and then mapping that $M_\mathrm{CO}$ back to $M_\mathrm{ZAMS}$ of a single-star model with the same $M_\mathrm{CO}$; we then adopt the yields corresponding to that single-star ZAMS mass. For example, for a binary star with $M_\mathrm{ZAMS}=60\,M_\odot$ and $M_\mathrm{CO}=14\,M_\odot$, this CO core mass corresponds to $M_\mathrm{ZAMS}=36\,M_\odot$ without the binary mass-transfer effect; we therefore approximate the binary-model yields by those of a single-star model with $M_{\mathrm{ZAMS}} = 36\,M_\odot$, and assume that the remaining mass is lost through winds. Although it has been argued that binary evolution can modify CCSN yields and thus affect GCE \citep[e.g.,][]{laplace2021Different,farmer2023Nucleosynthesis,kemp2025Binary}, in this work we use the same CCSN yield set \citep{chieffi2004Explosive} even when the explodability is taken from the binary-evolution cases.

\subsubsection{Model Ex-PUSH}\label{section:PUSH_explodability}

\citet{ebinger2019PUSHing,ebingerPUSHing2020} presented explodability as a function of $M_\mathrm{ZAMS}$ for three metallicities, $Z=0$, $10^{-4}Z_\odot$, and $Z_\odot$. In this study, we adopt their ``standard calibration" model.  We refer to the GCE model employing this explodability prescription as Ex-PUSH (following the titles of \citealt{ebinger2019PUSHing,ebingerPUSHing2020}). This model belongs to group Ex.

Because the explodability is provided at only three discrete metallicities, we interpolate it into the whole metallicity range in a manner similar to \citet{maltsevExplodability2025}. For $0\leq Z \leq 10^{-4}Z_\odot$, we connect the values linearly in $Z$. For $10^{-4}Z_\odot \leq Z \leq Z_\odot$, we connect the values such that the interpolation follows a dependence with slope proportional to $\log(Z/Z_\odot)$. For $Z \geq Z_\odot$, we adopt an extrapolation of the relation defined over $10^{-4}Z_\odot \leq Z \leq Z_\odot$. Moreover, we simply assume the stars with $M_\mathrm{ZAMS}\geq40M_\odot$ form BHs in model Ex-PUSH. With these choices, the explodability is summarized in the right panel of Figure~\ref{figure:explodability_schematic}.

\begin{figure*}[htb!]
    \includegraphics[width=18cm]{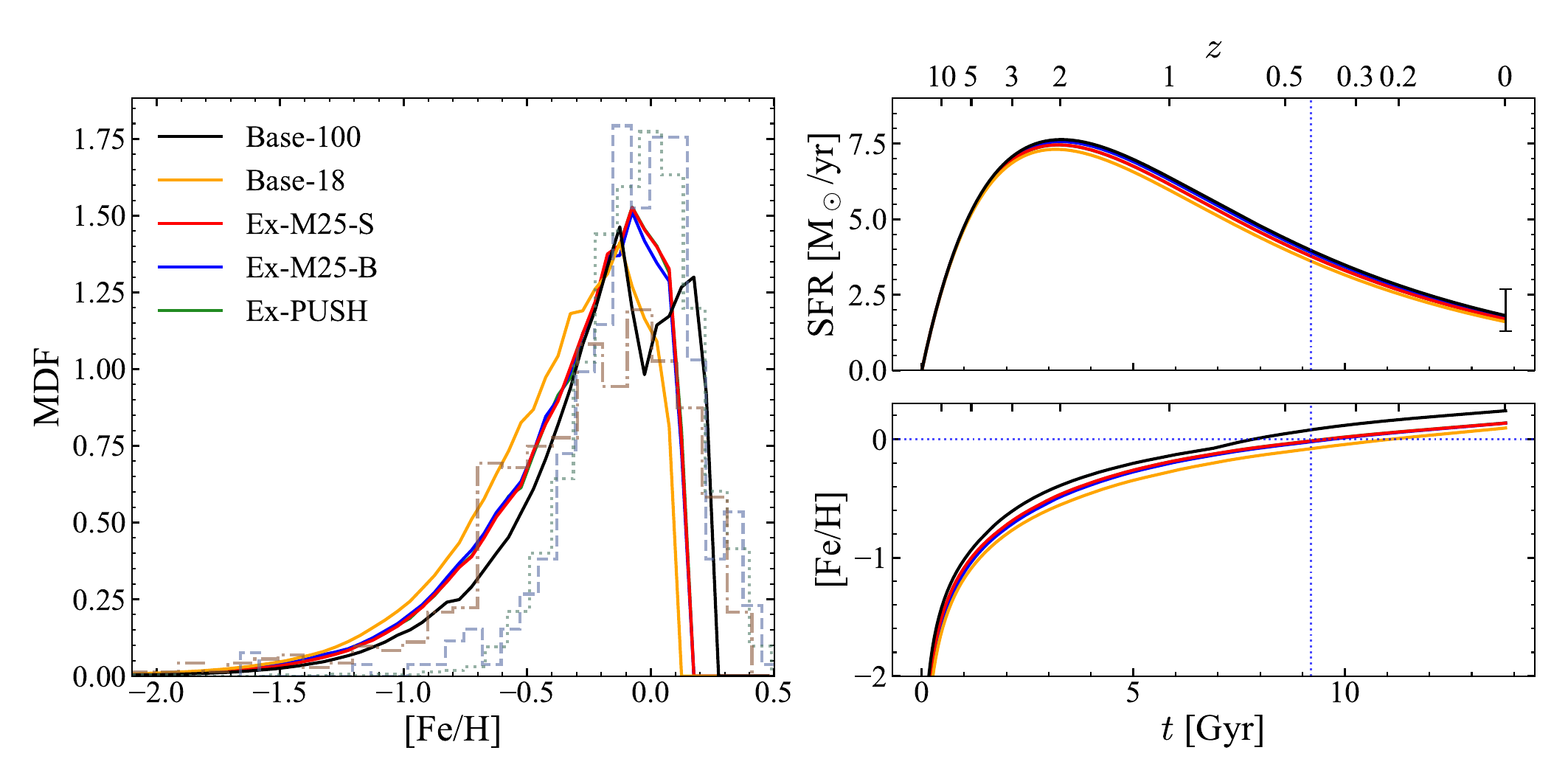}
\caption{
Metallicity distribution function (MDF) and the evolution of the star formation rate (SFR) and $[\mathrm{Fe/H}]$ are shown as indicators of model validation. The left panel displays the MDF as a function of $[\mathrm{Fe/H}]$, together with observational estimates from \citet{casagrandeNew2011} (blue dashed line), \citet{buder2019GALAH} (green dotted line), and \citet{bensby2014Exploring} (brown dash-dotted line). The right panels present the SFR and $[\mathrm{Fe/H}]$ as functions of cosmic time $t$ or redshift $z$. The blue dotted horizontal line in the bottom right panel denotes $[\mathrm{Fe/H}] = 0$, and the blue dotted vertical lines in the right panels mark $t = t_\odot$. The error bar in the top right panel indicates the observational present-day SFR \citep{elia2022Star}. Models are color-coded as follows: Base-100 (black), Base-18 (orange), Ex-M25-S (red), Ex-M25-B (blue), and Ex-PUSH (green).
}
    \label{figure:model_validity}
\end{figure*}


\subsection{Results}\label{sec:results}

\subsubsection{Model Validation}\label{section:model-validity}

Given a number of parameters involved, a GCE model must be calibrated to satisfy some basic observational constraints before discussing its outcome and implications. This subsection describes how our models are validated, using the MDF, as well as the time evolution of the SFR and $[\mathrm{Fe/H}]$\footnote{$[\mathrm{A/B}]$ denotes a number ratio relative to the solar value, defined as $\log(N_\mathrm{A}/N_\mathrm{B})-\log(N_\mathrm{A,\odot}/N_\mathrm{B,\odot})$.}.

The MDF is widely used as a validation of GCE models. As shown in Figure~\ref{figure:model_validity}, the MDFs computed in all models in group Ex are consistent with the observational MDF.

In GCE studies, $[\mathrm{Fe/H}]$ is often used as a practical evolutionary coordinate, rather than the cosmic time $t$, because many observational relations are discussed as a function of $[\mathrm{Fe/H}]$. Accordingly, the model should reproduce the expectation that $[\mathrm{Fe/H}]\approx 0.0$ at $t=t_\odot=9.2\,\mathrm{Gyr}$. This constraint guaranties a correct temporal sequence at $[\mathrm{Fe/H}] = 0.0$. Figure~\ref{figure:model_validity} shows that all models in group Ex pass through $[\mathrm{Fe/H}]\approx 0.0$ at $t=t_\odot$. In addition, all models reproduce the present-day SFR inferred from observations, $2.0_{-0.7}^{+0.7}\,M_\odot\,\mathrm{yr}^{-1}$ \citep{elia2022Star}.

Although small model-to-model variations are present, the overall evolution of $[\mathrm{Fe/H}]$ does not differ substantially among the models. Unlike oxygen, Fe from CCSNe is supplied mainly by relatively low-mass massive stars (see Appendices \ref{sec:comp_yields} and \ref{sec:yield_O_Fe}). As a result, changes in whether stars with $M_\mathrm{ZAMS}\gtrsim 18\,M_\odot$ explode as CCSNe or collapse to BHs have only a limited impact on the global Fe enrichment. For these reasons, the predicted evolution of $[\mathrm{Fe/H}]$ remains broadly similar across the different models considered here.

\begin{figure*}[htb!]
    \includegraphics[width=18cm]{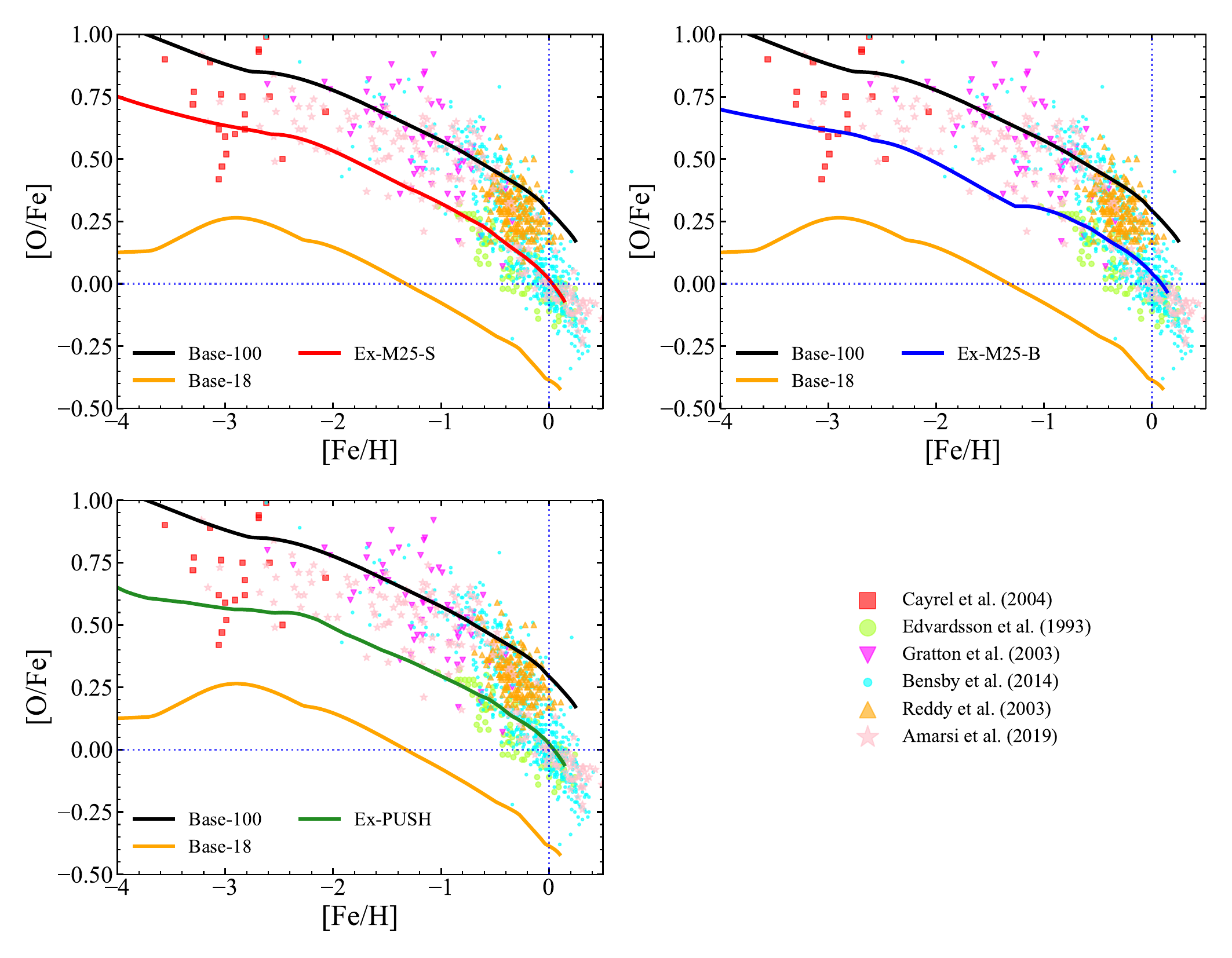}
\caption{
Evolution of $[\mathrm{O/Fe}]$ as a function of $[\mathrm{Fe/H}]$ for group Ex. The top-left, top-right, and bottom-left panels show the results for models Ex-M25-S, Ex-M25-B, and Ex-PUSH, respectively. All panels also display the reference models Base-100 and Base-18. The colored symbols represent observational data; red squares from \citet{cayrel2004First}, green-yellow circles from \citet{edvardsson1993Chemical}, magenta downward triangles from \citet{gratton2003Abundances}, cyan points from \citet{bensby2014Exploring}, orange upward triangles from \citet{reddy2003Chemical}, and pink stars from \citet{amarsi2019Carbon}. The blue dotted vertical and horizontal lines denote $[\mathrm{Fe/H}]=0$ and $[\mathrm{O/Fe}]=0$, respectively. Colors of the models are the same as in Figure~\ref{figure:model_validity}.
}
    \label{figure:main_O_Fe}
\end{figure*}

\begin{figure}[htb!]
    \hspace{-1cm}
    \includegraphics[width=10cm]{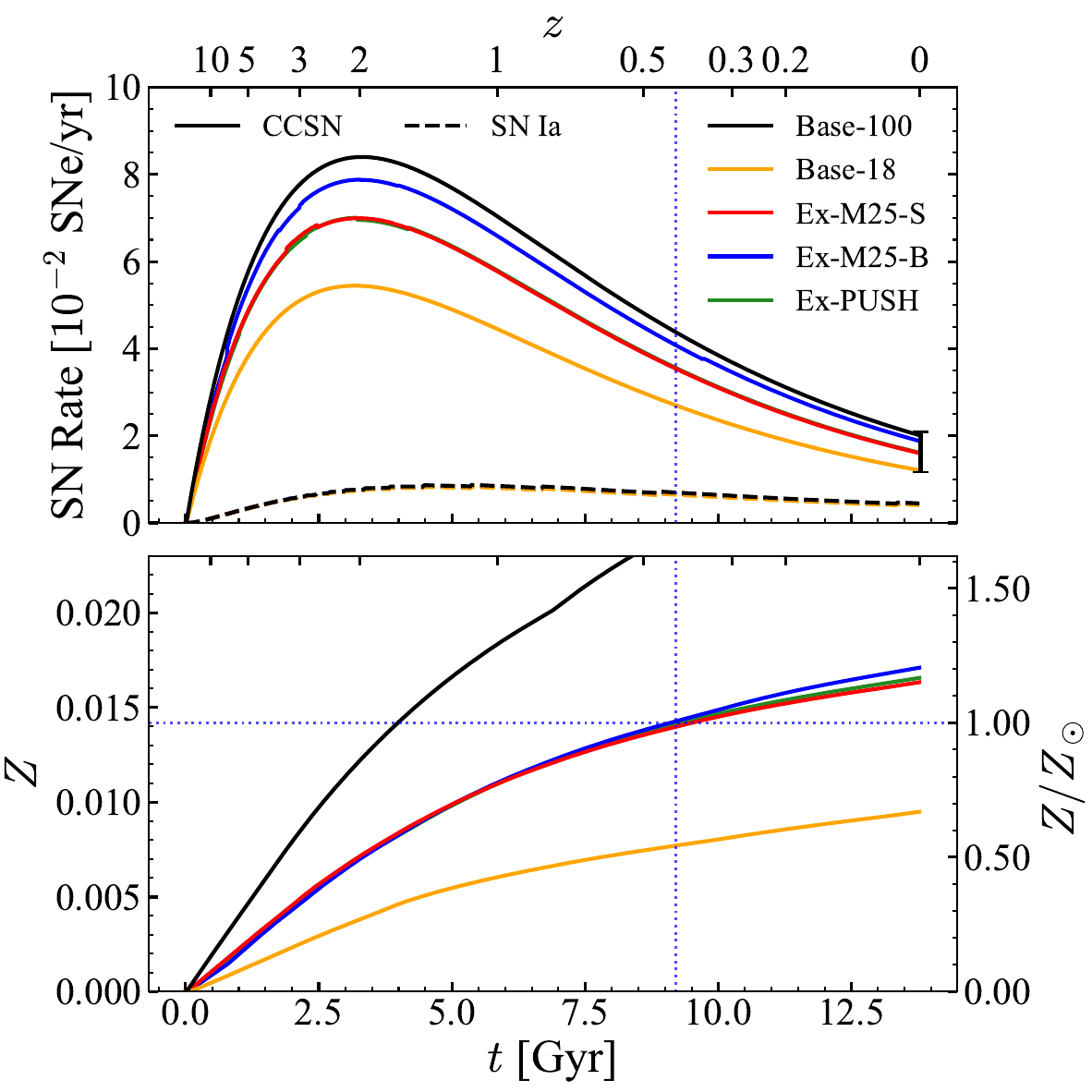}
\caption{
Evolution of metallicity and supernova (SN) rate. The top panel shows the evolution of CCSN rate (solid lines) and SN Ia rate (dashed lines). The bottom panel shows the evolution of metallicity $Z$. The blue dotted horizontal line in the bottom line denotes $Z=Z_\odot$, and the blue dotted vertical lines mark $t = t_\odot = 9.2\,\mathrm{Gyr}$. The error bar in the top right panel indicates the observational present-day CCSN rate \citep{rozwadowska2021Rate}. Colors are the same as in Figure~\ref{figure:model_validity}.
}
    \label{figure:Metallicity_SNR}
\end{figure}

\subsubsection{Group Ex}\label{section:Explodability-Results}

Figure~\ref{figure:main_O_Fe} shows the $[\mathrm{O/Fe}]$--$[\mathrm{Fe/H}]$ relations obtained with the group Ex models. Models Base-100 and Base-18 are included here only as comparison baselines and are not tuned to provide a good match. As pointed out by \citet{suzukiConstraining2018}, the model Base-18, in which stars above $18\,M_\odot$ do not explode, fails to reproduce the observed trend. In contrast, the group Ex models reproduce the observations both at low and at high $[\mathrm{Fe/H}]$. Moreover, all group Ex models satisfy $[\mathrm{O/Fe}]\approx 0.0$ at $[\mathrm{Fe/H}]=0.0$, indicating that they reproduce the solar abundance ratio. The observed behavior at super solar metallicity ($[\mathrm{Fe/H}]\gtrsim 0$) may also be reproduced, as implied by the slope around $[\mathrm{Fe/H}]\approx 0$.

Figure~\ref{figure:Metallicity_SNR} presents the time evolution of the SN rate and the metallicity for the group Ex models. The time evolution of the metallicity shows that all group Ex models satisfy $Z\approx Z_\odot$ at $t=t_\odot$. In Section~\ref{section:model-validity}, we noted that the SFR does not vary strongly among our models, whereas the SN rate can differ substantially due to their explodability. In particular, although models Ex-M25-S and Ex-M25-B show nearly identical chemical evolution and SFR in Figures~\ref{figure:model_validity} and \ref{figure:main_O_Fe}, the difference in their SN rates is not negligible (Figure~\ref{figure:Metallicity_SNR}). As discussed in Section~\ref{section:M25_explodability}, this arises because in the binary case the removal of the outer envelope reduces $M_\mathrm{CO}$ relevant for nucleosynthesis; even if a star with large $M_\mathrm{ZAMS}$ explodes, its yields are similar to those of a lower-mass single star, with a reduced oxygen production compared to an unstripped explosion \citep{schneider2023Bimodal}. Consequently, the number of CCSNe can increase while the total oxygen enrichment changes only slightly.

\begin{figure}[htb!]
    \hspace{-0.5cm}
    \includegraphics[width=9cm]{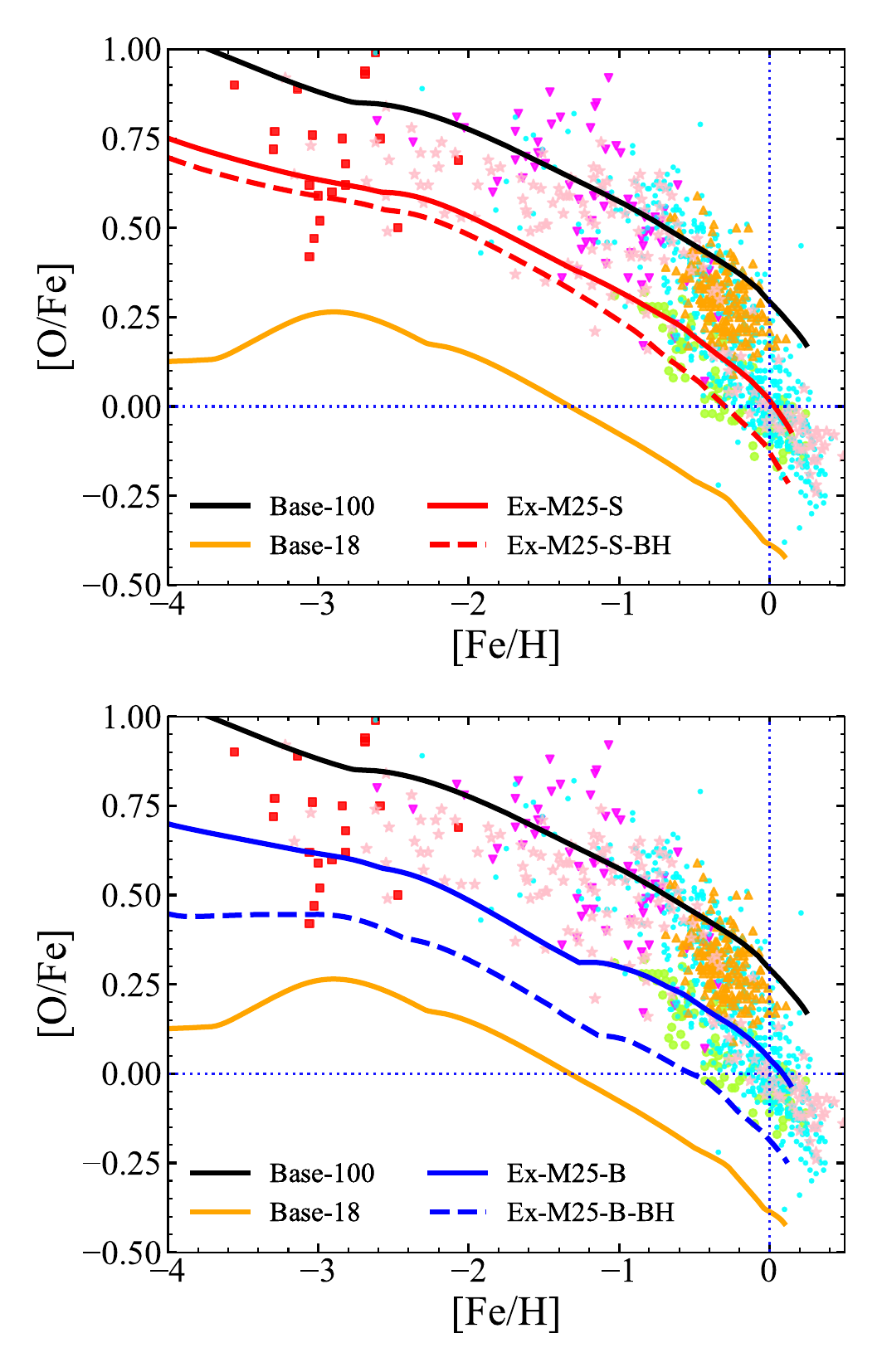}
\caption{
Evolution of $[\mathrm{O/Fe}]$ for group Ex-BH. The solid lines correspond to the same models shown in Figure~\ref{figure:main_O_Fe} (models Base-100, Base-18, Ex-M25-S, and Ex-M25-B), while the dashed lines show the evolution of group Ex-BH (see Figure~\ref{figure:explodability_schematic} and Section~\ref{section:M25_explodability}). The blue vertical and horizontal reference lines, as well as the observational data points, are identical to those in Figure~\ref{figure:main_O_Fe}.
}
    \label{figure:strict_O_Fe}
\end{figure}

\subsubsection{Group Ex-BH}\label{section:Strict-Explodability-Results}

Figure~\ref{figure:strict_O_Fe} shows the $[\mathrm{O/Fe}]$--$[\mathrm{Fe/H}]$ relations obtained with the group Ex-BH models. Model Ex-M25-S-BH reproduces the observed trend at low $[\mathrm{Fe/H}]$, but behaves significantly lower $[\mathrm{O/Fe}]$ at high $[\mathrm{Fe/H}]$. This behavior arises because, at $Z\approx 0$, relatively massive stars ($20\,M_\odot \lesssim M_\mathrm{ZAMS}\lesssim 30\,M_\odot$) still explode, whereas at high metallicity most high-mass massive stars fail to explode, leading to the rapid reduction in $[\mathrm{O/Fe}]$. This implies that a larger fraction of oxygen-providing massive stars have to explode at high metallicity than postulated in model Ex-M25-S-BH. Model Ex-M25-B-BH predicts $[\mathrm{O/Fe}]$ values that are systematically below the observations over the full range of $[\mathrm{Fe/H}]$. In model Ex-M25-B-BH, almost all high-mass massive stars form BHs, and the resulting oxygen production is therefore reduced globally.

At $[\mathrm{Fe/H}]=0$, the group Ex-BH models differ in their $[\mathrm{O/Fe}]$ from the solar value by more than $0.1$ dex, and it is difficult to reduce this gap to $[\mathrm{O/Fe}]\approx0$ by tuning parameters (most notably $\epsilon_\mathrm{out}$) while still maintaining reasonable agreement with other observations (see Section~\ref{section:results_simplified} for details). In addition, the IMF-averaged CCSN yields of the Ex-BH models at solar metallicity provide too small $[\mathrm{O/Fe}]$, close to zero. Since the inclusion of SNe~Ia lowers $[\mathrm{O/Fe}]$ substantially, achieving $[\mathrm{O/Fe}]=0$ at $[\mathrm{Fe/H}]=0$ is generally difficult for such a model under typical assumptions (see Appendix~\ref{sec:yield_O_Fe} for detail).

Improvement of $[\mathrm{O/Fe}]$ at low metallicity is also difficult. Similarly to the situation at solar matallicity, a key is the number ratio of O-producing CCSNe (determined by the explodability at low $Z$) to Fe-producing CCSNe (see Appendix~\ref{sec:yield_O_Fe}). It is thus difficult to adjust the offset of $[\mathrm{O/Fe}]$ at low metallicity by simply adjusting other GCE parameters.

In model Ex-M25-S-BH, progenitors with $M_\mathrm{ZAMS}\gtrsim 18\,M_\odot$ predominantly form BHs, making this model a useful limiting case for exploring the possible relation between the RSG problem and the CCSN explodability through the GCE arguments. The resulting $[\mathrm{O/Fe}]$ evolution in group Ex-BH indicates that reproducing the [O/Fe] evolution under such a strong suppression of high-mass explosions at high $[\mathrm{Fe/H}]$ is difficult (see above). Based on these results in this section, we further discuss a possible solution of the RSG problem in the next section.

\begin{figure*}[htb!]
    \includegraphics[width=18cm]{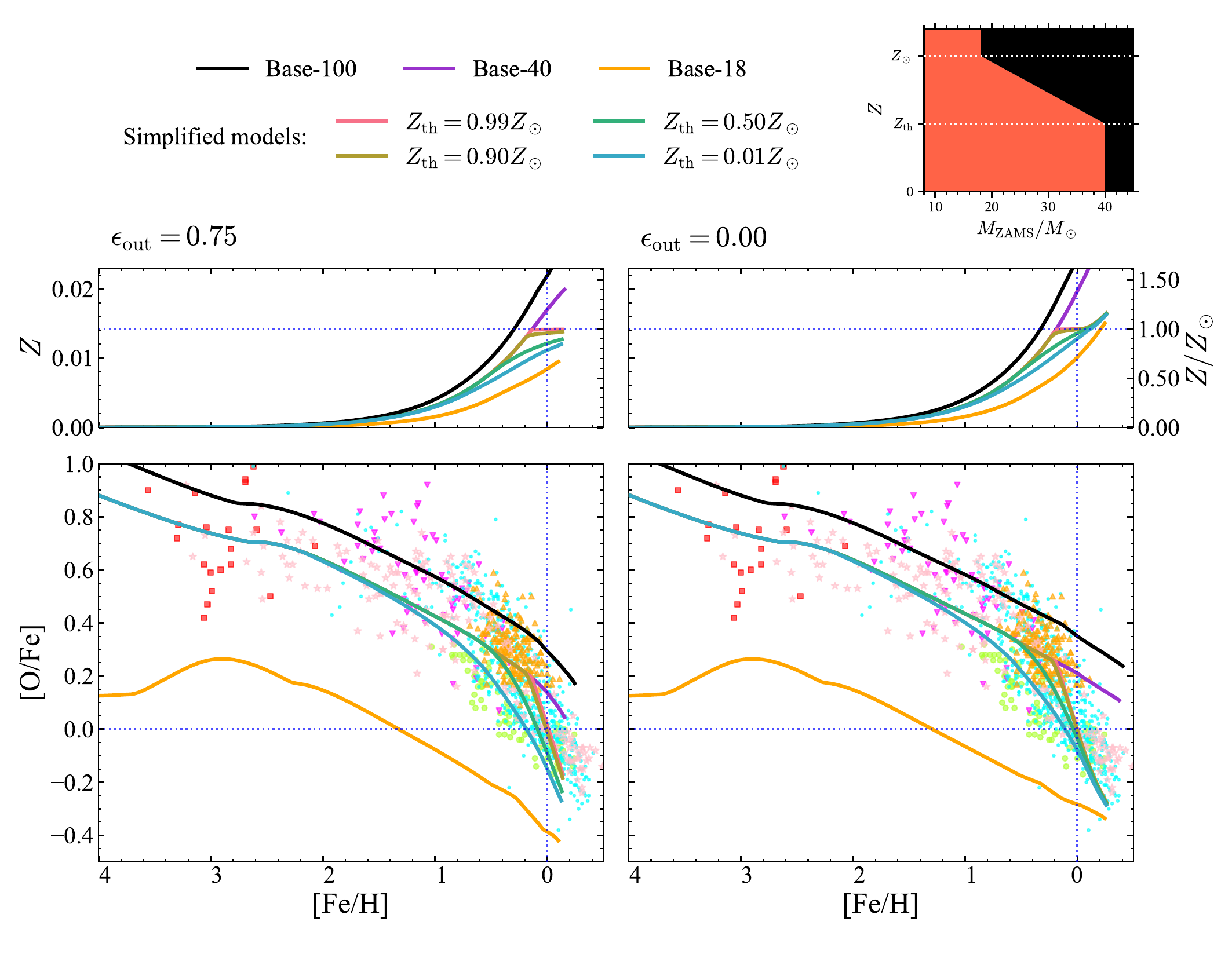}
\caption{
Evolution and explodability in the simplified models. The top-right sub-panel presents the explodability as a function of ZAMS mass (vertical axis) and metallicity (horizontal axis), where $Z_\mathrm{th}$ is the threshold metallicity treated as a free parameter in Eq.~\eqref{eq:Mth}. The main top panels show the $[\mathrm{Fe/H}]$--metallicity planes, and the bottom panels display the $[\mathrm{O/Fe}]$--$[\mathrm{Fe/H}]$ planes. The left and right columns correspond to outflow efficiencies of $\epsilon_\mathrm{out}=0.75$ and $\epsilon_\mathrm{out}=0.00$, respectively. The colored lines represent simplified models with $Z_\mathrm{th} = 0.99Z_\odot$ (red), $0.90Z_\odot$ (yellow), $0.50Z_\odot$ (green), and $0.01Z_\odot$ (blue). The models Base-100 (black), Base-40 (purple), and Base-18 (orange) are shown for reference. Blue dotted lines indicate $[\mathrm{Fe/H}]=0$, $[\mathrm{O/Fe}]=0$, and the solar metallicity ($Z = Z_\odot$). Observational data are the same as those in Figure~\ref{figure:main_O_Fe}.
}
    \label{figure:simplified_models}
\end{figure*}
    

\section{Simplified models}\label{sec:simplified_model}


\subsection{Numerical Setting}\label{section:setting_simplified}

To deepen our understanding of the effects of metallicity-dependent explodability on GCE, and to examine whether such a dependence can potentially account for the RSG problem, we introduce a set of simplified models for the explodability in this section. Instead of those motivated by the stellar-evolution and SN-explosion studies as examined in the previous section, here we assume a minimal form of metallicity-dependent explodability in which CCSNe occur up to $40\,M_\odot$ at low metallicity (i.e., at low $[\mathrm{Fe/H}]$). This choice is motivated by the result from group Ex-BH models (Section~\ref{section:Strict-Explodability-Results} and Figure~\ref{figure:strict_O_Fe}); if the explosions of massive stars providing oxygen are reduced at low metallicity, the predicted $[\mathrm{O/Fe}]$ cannot reproduce the observational trends. At $Z\approx Z_\odot$, on the other hand, we require that the maximum exploding mass is $\sim 18\,M_\odot$, i.e., one possible solution to the RSG problem. These two regimes are then connected smoothly. We note that the RSG problem is primarily an issue in the local universe, and thus no constraint is placed on the nature of SN progenitors at low metallicity. 

We define $Z_\mathrm{th}$ as the maximum metallicity at which CCSNe occur up to $40\,M_\odot$. Using $Z_\mathrm{th}$, we define the upper boundary of the CCSN-producing mass range ($M_\mathrm{th}$) in a simplified prescription, given as
\begin{align}
    M_\mathrm{th} =
    \begin{cases}
    40\,M_\odot, & \textrm{for\ } Z < Z_\mathrm{th},\\[4pt]
    \left[40 + (18-40)\dfrac{Z-Z_\mathrm{th}}{Z_\odot-Z_\mathrm{th}}\right]M_\odot, & \textrm{for\ } Z_\mathrm{th}\le Z < Z_\odot,\\[4pt]
    18\,M_\odot, & \textrm{for\ } Z \ge Z_\odot.
    \end{cases}
    \label{eq:Mth}
\end{align}
The top-right sub-panel of Figure~\ref{figure:simplified_models} shows the resulting explodability in this simplified model. We consider $Z_\mathrm{th}=0.01Z_\odot$, $0.50Z_\odot$, $0.90Z_\odot$, and $0.99Z_\odot$.

As a reference, we also include model Base-40, in which the exploding mass range is assumed to be $8\textrm{--}40\,M_\odot$ at all metallicities. Unless otherwise stated, we adopt the same parameters as in Section~\ref{section:main-method}. In addition, we test the impact of the outflow strength by considering not only $\epsilon_\mathrm{out}=0.75$ (our fiducial choice in Section~\ref{section:main-method}) but also a case with no outflow, $\epsilon_\mathrm{out}=0.00$. For $\epsilon_\mathrm{out}=0.00$, we set the initial inflow rate to $\dot{M}_\mathrm{in,0}=8\,M_\odot\,\mathrm{yr}^{-1}$ in order to match the observational SN rate and the observational SFR. The effects on the MDF and the time evolution of $[\mathrm{Fe/H}]$ are described in Section~\ref{section:results_simplified}. The settings of the simplified models are also summarized in Table~\ref{table:GCEparams}.


\subsection{Results}\label{section:results_simplified}

\subsubsection{Reference model ($\epsilon_\mathrm{out}=0.75$)}

Figure~\ref{figure:simplified_models} shows the $[\mathrm{O/Fe}]$--$[\mathrm{Fe/H}]$ relations obtained with the simplified models. We first describe the results for $\epsilon_\mathrm{out}=0.75$ to examine the basic behavior.

For $Z_\mathrm{th}=0.01Z_\odot$, the predicted track deviates from the observations and passes below the solar-abundance point. In this case, the metallicity at $[\mathrm{Fe/H}]=0$ is $Z=0.0112$, which is smaller than $Z_\odot$.

For $Z_\mathrm{th}=0.50Z_\odot$, the model does not deviate strongly from the observational trend over the main range, but the decline of $[\mathrm{O/Fe}]$ is steep, indicating that the model may depart significantly at super solar metallicity. The metallicity at $[\mathrm{Fe/H}]=0$ is $Z=0.0121$, only slightly below $Z_\odot$. However, Eq.~\eqref{eq:Mth} yields $M_\mathrm{th}=24.4\,M_\odot$ at $Z=0.0121$, which remains far above $18\,M_\odot$ and therefore does not resolve the RSG problem. Nevertheless, because the metallicity at $z=0$ (present day), in which the RSG problem has been studied, becomes somewhat larger, $M_\mathrm{th}$ for the progenitors of local SNe IIP could become smaller as well, and the situation may be slightly improved.

For $Z_\mathrm{th}=0.90Z_\odot$ and $0.99Z_\odot$, both the $[\mathrm{O/Fe}]$--$[\mathrm{Fe/H}]$ relation and the metallicity evolution are close to the observational constraints. However, these cases effectively require a very rapid change in explodability near $Z\approx Z_\odot$, which may appear chance coincidence that might require fine-tuning of the stellar evolution physics. A comparison between  Base-40 and the case $Z_\mathrm{th}=0.99Z_\odot$ shows that, once $Z$ reaches $Z_\mathrm{th}$, the metallicity in the $Z_\mathrm{th}=0.99Z_\odot$ model evolves very little and becomes nearly constant. Qualitatively, this can be understood as follows: CCSNe enrich the gas and increase $Z$, which reduces $M_\mathrm{th}$; a smaller $M_\mathrm{th}$ suppresses the metal supply, acting as a restoring feedback on $Z$ and making further metallicity growth inefficient.  In this sense, the system in this model is self-regulating -- the present-day metallicity is largely controlled by $Z_\mathrm{th}$, i.e., by the stellar physics. This may indeed be an interesting possibility, but the characteristic plateau in the metallicity evolution may violate some observational constraints. Furthermore, this self-regulating behavior may well be smoothed out in more realistic GCE models beyond the one-box treatment. 

Overall, the results of our simplified models for $\epsilon_\mathrm{out}=0.75$ imply the following two points. First, these models exhibit a rapid decrease in $[\mathrm{O/Fe}]$ once the system enters the regime $Z>Z_\mathrm{th}$. This can be attributed to the fact that the outflow removes oxygen from the system, thereby allowing the abundance pattern to immediately reflect the IMF-averaged CCSN yields at given metallicity (see Appendix~\ref{sec:yield_O_Fe} for detail). This drives the rapid decrease in $[\mathrm{O/Fe}]$ toward the track of model Base-18. Indeed, compared to model Base-40, this effect of decreasing [O/Fe] toward the solar metallically could help to reproduce the [O/Fe] evolution in the range above [Fe/H] $\sim 0.1$; this is an interesting effect as this trend is basically reproduced by the increasing SN Ia contribution in most of GCE models, but the present work shows that the CCSN explodability could also contribute to this behavior. 

Second, as $Z_\mathrm{th}$ increases, the predicted $[\mathrm{O/Fe}]$ at $[\mathrm{Fe/H}]=0$ and the metallicity at $t=t_\odot$ rise, eventually approximately matching the solar values at $Z_\mathrm{th} \approx Z_\odot$. More generally, this behavior can be understood by considering the composition of the metal. Given that oxygen and iron abundance predominate the total metal mass, the condition of simultaneously satisfying $Z=Z_\odot$ and $[\mathrm{Fe/H}]=0$ requires $[\mathrm{O/Fe}]\approx0$. In our simplified model, increasing $Z_\mathrm{th}$ raises the metallicity at $[\mathrm{Fe/H}]=0$, but achieving $Z \approx Z_\odot$ at $[\mathrm{Fe/H}]=0$ becomes possible only when $Z_\mathrm{th}\approx Z_\odot$. These two facts explains the behavior of both the predicted $[\mathrm{O/Fe}]$ and metallicity at $[\mathrm{Fe/H}]=0$. Moreover, when considering alternative simplified-model setups, the self-regulation mechanism discussed above is expected to drive the system toward a state effectively pinned near $Z_\mathrm{th}\approx Z_\odot$ as $Z_\mathrm{th}$ increases. Hence, the same qualitative trend should hold generically, largely independent of the specific choices for the other model parameters.

\subsubsection{No outflow model ($\epsilon_\mathrm{out}=0.00$)}

To reconcile the deficiency of oxygen, another possibility is to reduce the outflow so that oxygen produced by CCSNe is efficiently retained within the Galaxy.  Therefore, we next consider the case with no outflow, $\epsilon_\mathrm{out}=0.00$, as an extreme case. Compared to $\epsilon_\mathrm{out}=0.75$, Figure~\ref{figure:simplified_models} shows the slope of $[\mathrm{O/Fe}]$ around $[\mathrm{Fe/H}]\approx 0$ becomes somewhat shallower, and the metallicity continues to increase even after entering the regime $Z>Z_\mathrm{th}$, rather than saturating abruptly. This is because, without outflow, previously produced metals are not removed from the system, so the evolution of $[\mathrm{O/Fe}]$ proceeds more gradually than in $\epsilon_\mathrm{out}=0.75$, as expected. Moreover, retaining metals within the system allows $[\mathrm{O/Fe}]$ to exceed the IMF-averaged CCSN value at solar metallicity, which is otherwise difficult to achieve (see Appendix~\ref{sec:yield_O_Fe}). In particular, for $Z_\mathrm{th}=0.50Z_\odot$, the $\epsilon_\mathrm{out}=0.75$ case suffered from insufficient metallicity at $[\mathrm{Fe/H}]=0$, which in turn made $M_\mathrm{th}$ too large. In contrast, for $\epsilon_\mathrm{out}=0.00$ we obtain $Z=0.0136$ at $[\mathrm{Fe/H}]=0$, corresponding to $M_\mathrm{th}=19.9\,M_\odot$, which is not far from the target value of $\sim 18\,M_\odot$. The slope at super solar metallicity is also favorable. However, for $Z_\mathrm{th}=0.01\,Z_\odot$, the predicted $[\mathrm{O/Fe}]$ at $[\mathrm{Fe/H}]=0$ remains slightly below zero. A comparison of model Base-18 with $\epsilon_\mathrm{out}=0.75$ and $0.00$ shows that even removing outflows increases $[\mathrm{O/Fe}]$ at $[\mathrm{Fe/H}]=0$ by only $\sim0.1$ dex. Since the model with $Z_\mathrm{th}=0.01\,Z_\odot$ already yields $[\mathrm{O/Fe}]\approx-0.15$ for $\epsilon_\mathrm{out}=0.75$ at $Z_\mathrm{th}=0.01\,Z_\odot$, adopting $\epsilon_\mathrm{out}=0.00$ is still insufficient to reach the solar ratio.

From these results, we infer that models with $\epsilon_\mathrm{out}=0.00$ and $Z_\mathrm{th}\gtrsim 0.50Z_\odot$ may be able to reproduce various observations (standard GCE calibrations and the [O/Fe] evolution) reasonably well, while potentially alleviating the RSG problem. We note that there are a few remaining issues; at $t=t_\odot$ the model yields $[\mathrm{Fe/H}]=0.082$, which is somewhat high; the peak of MDF is slightly larger than the value inferred from the observations. However, we do not regard this as a major discrepancy for the present qualitative exploration. Although a model without outflow may appear unrealistic, we adopt $\epsilon_\mathrm{out}=0.00$ only as an extreme limiting case. We note that models with sufficiently small outflow efficiencies, $\epsilon_\mathrm{out}\lesssim 0.20$, remain broadly compatible with the observational constraints. Moreover, the apparent absence/weakness of the outflows in a one-zone model could be partly alleviated in a multizone framework. For example, scenarios in which metal-rich stars migrate into the solar neighborhood have been discussed in the context of GCE \citep{tsujimotoGalactic2022}. Such migration could effectively compensate for the reduction in the local metal budget caused by outflows, because metal rich stars formed in the inner Galaxy may subsequently migrate into the solar neighborhood; in this sense, the net effect of radial migration can resemble that of a metal bearing inflow, potentially leading to chemical evolution trends similar to those of a one zone model without outflows even when outflows are present.

\subsubsection{Summary}
In summary, the exercise here with the simplified models yields several interesting insights. With the presence of strong outflow from the system frequently assumed in GCE models, it is difficult to reconcile the RSG problem by the metallicity-dependent explodability. Reducing the outflow allows previously produced metals to remain in the system, making it possible to reach higher metallicity even for $Z_\mathrm{th}=0.50Z_\odot$, thereby improving both the match to the GCE observations and $M_\mathrm{th}$ which can solve the RSG problem. In the end, while weakening outflow can increase the likelihood of reproducing the solar composition, the simplified models still indicate that the transition metallicity must be close to solar, i.e., $Z_\mathrm{th}\sim Z_\odot$, to connect to $M_\mathrm{th}=18\,M_\odot$ at $Z=Z_\odot$. 

Although the physical origin of such a strongly metallicity-dependent explodability should ultimately be examined with detailed SN explosion models, such an investigation is beyond the scope of the present work. Observationally, however, the number of CCSNe per SFR has been suggested to decline sharply over a relatively narrow range of metallicity \citep{pessiMetallicityDependenceOccurrence2023}, which may point to a strong metallicity dependence in CCSN production similar to that explored here as a potential way to alleviate the RSG problem.


\section{Discussion}\label{sec:discussion}
In this section, we discuss implications of our results from several perspectives, and perform additional validation tests to assess the robustness of our conclusions. In Section~\ref{section:const_explodability}, we argue that GCE provides meaningful constraints on metallicity-dependent explodability. In Section~\ref{section:alpha_elements}, we examine how changes in explodability affect the GCE of $\alpha$-elements other than oxygen. Related to this, Section~\ref{section:sub-CH} investigates the impact of including sub-Ch SN~Ia on GCE. In Section~\ref{section:BH_yields}, we discuss the possible impact of including yields from BH-forming events, as suggested by recent studies. In Section~\ref{section:multi-infall}, we discuss the implications of multi-infall GCE models. In Section~\ref{section:MD-IMF}, we apply a recently proposed metallicity-dependent IMF to our models. Finally, in Section~\ref{section:missing-Methods}, we discuss how metallicity-dependent explodability modifies the cosmic CCSN rate density and whether it can help alleviate the missing SN problem.


\subsection{Constraining Explodability with Galactic Chemical Evolution}\label{section:const_explodability}

In Sections~\ref{sec:methods} and \ref{sec:simplified_model}, we demonstrate that not all forms of explodability can reproduce the observed GCE. This implies that GCE can impose constraints on the allowed forms of metallicity-dependent explodability. Moreover, within these constraints, Section~\ref{sec:simplified_model} shows that it is, in principle, possible to construct metallicity-dependent explodability models that may alleviate the RSG problem. Namely, if the high-mass stars with $M_{\rm ZAMS} \gtrsim 18\,M_\odot$ would fail to explode at solar metallicity but explode at low metallicity to provide an early enrichment of oxygen, this scenario may simultaneously explain both the RSG problem and the [O/Fe] evolution in GCE. For this scenario to work, the transition of the explodability should be at sub-solar metallicity ($Z \gtrsim 0.5 Z_\odot$). The idea is broadly consistent with suggestions that relatively massive ($M_{\rm ZAMS}=17$--$25\,M_\odot$) progenitors can produce Type~II SNe at low metallicity \citep{andersonLowestmetallicity2018}, although further observational constraints on progenitor masses and SN demographics in metal-poor environments are needed.

Attempts to constrain explodability with GCE have been discussed in other studies. \citet{jost2025Neutrinodriven} argued that reproducing GCE requires at least $\sim 50\%$ of massive stars to explode as CCSNe, but this requirement inevitably depends on the assumed CCSN-producing mass range, also as a function of metallicity. For example, if all stars above $18\,M_\odot$ are assumed to fail, then roughly two-thirds of massive stars ($M_\mathrm{ZAMS}\gtrsim 8\,M_\odot$) are still to explode, yet this is insufficient to reproduce the oxygen evolution, as emphasized by \citet{suzukiConstraining2018}. Since models Ex-M25-S-BH and Ex-M25-B-BH struggle to reproduce the observations despite having a broader mass range for CCSNe than this scenario, an even larger CCSN fraction is likely required. Therefore, comparing our results with \citet{jost2025Neutrinodriven} implies that GCE constraints on explodability are governed not simply by the total CCSN fraction, but also critically by which progenitor-mass ranges contribute to successful explosions and chemical enrichment.

While the present study highlights a possible relation of the RSG problem with the explodability and GCE, the power of the GCE arguments to constrain the nature of explodability as we suggest here applies even if the RSG problem would be treated separately; for example, another suggestion to remedy the RSG problem is that massive stars with $M_{\rm ZAMS} \gtrsim 18 M_\odot$ would indeed explode but as an SN type different from SNe IIP, e.g., turning into Wolf-Rayet stars to explode as SNe Ib/Ic \citep{smarttProgenitors2009, suzukiConstraining2018}. Still, at least a fraction of such stars must form a BH with little ejection of metal-rich materials, to explain abundant existence of binary BHs as probed by gravitational waves \citep{abbott2023PhRvX..13a1048A}. On the other hand, the present work shows that some of them must also explode and eject oxygen into ISM \citep[see also][]{suzukiConstraining2018}, and there are indeed some SNe that are suggested to be explosions of stars with $M_{\rm ZAMS} \gtrsim 18 M_\odot$ \citep[e.g.,][]{maeda2026PASJ...78L...1M}. Our understanding of the stellar evolution and explodability is still incomplete, and the present model suggests that the consistency with GCE must be considered when these issues are tackled. 

\begin{figure*}[htb!]
    \includegraphics[width=18cm]{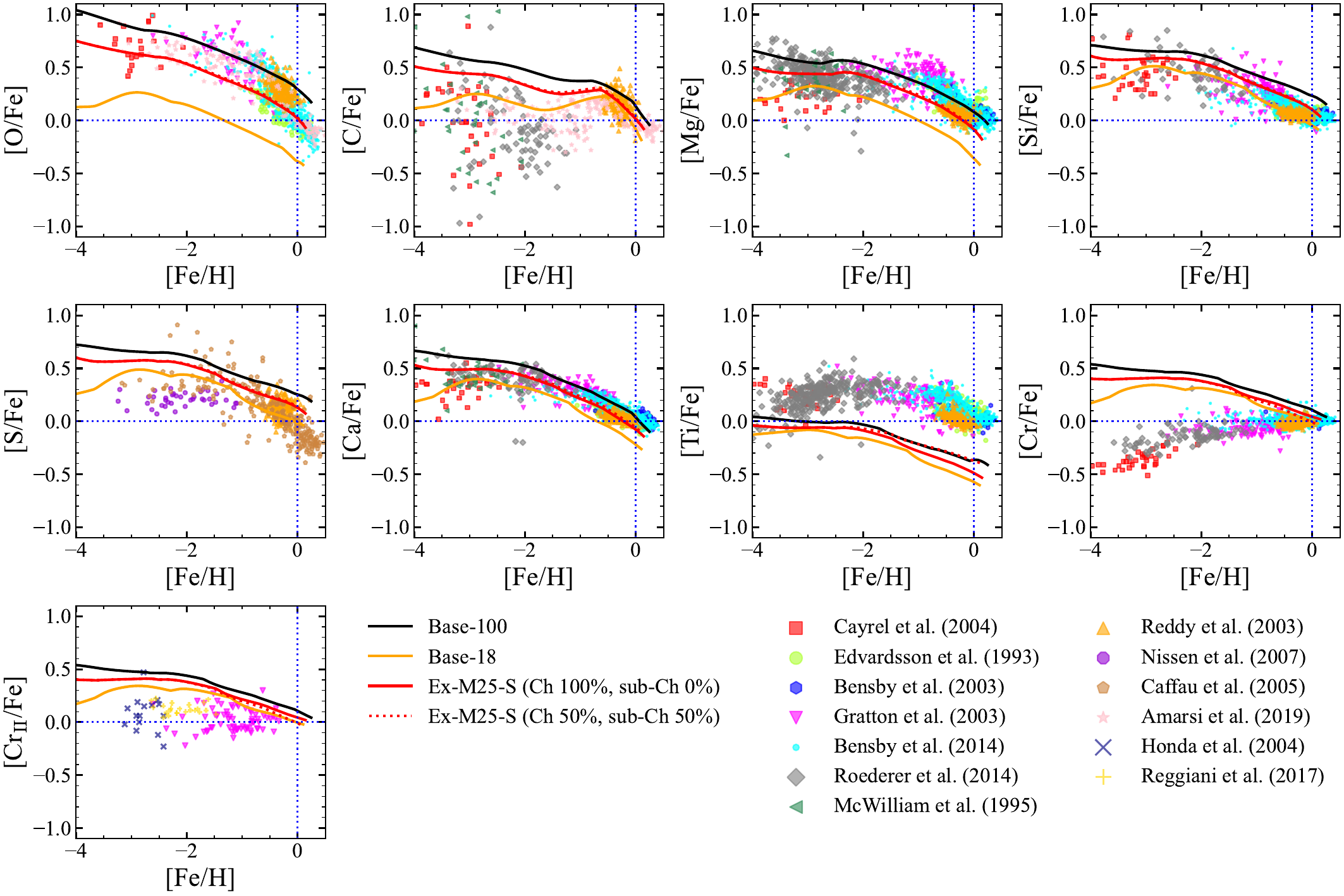}
\caption{
Comparison of the evolutionary trends of oxygen and other $\alpha$-elements (C, Mg, Si, S, Ca, Ti, and Cr) as a function of $[\mathrm{Fe/H}]$. The colors and line styles for models Base-100, Base-18, and Ex-M25-S are the same as in Figure~\ref{figure:main_O_Fe}. The dotted red line represents the model Ex-M25-S in which 50\% of all SNe~Ia are assumed to be sub-Ch~SNe~Ia, whereas the standard model Ex-M25-S assumes 100\% Ch~SNe~Ia. The colored symbols are the same as in Figure~\ref{figure:main_O_Fe}, with additional datasets shown as blue hexagons from \citet{bensby2003Elemental}, gray diamonds from \citet{roederer2014SEARCH}, green left triangles from \citet{mcwilliam1995Spectroscopic}, violet octagons from \citet{nissen2007Sulphur}, brown pentagons from \citet{caffau2005Sulphur}, navy crosses from \citet{honda2004Spectroscopic}, and yellow plus signs from \citet{reggiani2017Constraining}. For $[\mathrm{Cr\,\text{\sc ii}/Fe}]$, we include only observational data explicitly identified as Cr\,{\sc ii} in the original literature. The blue dotted vertical and horizontal lines indicate $[\mathrm{Fe/H}] = 0$ and $[\mathrm{O/Fe}] = 0$, respectively.
}
    \label{figure:alpha_elements}
\end{figure*}


\subsection{$\alpha$-elements}\label{section:alpha_elements}

We now examine how changes in explodability affect the evolution of $\alpha$-elements other than oxygen: C, Mg, Si, S, Ca, Ti, and Cr. Figure~\ref{figure:alpha_elements} shows the evolution of these elements (including oxygen) for models Base-100, Base-18, and Ex-M25-S. All elements exhibit the standard qualitative behavior, i.e., a plateau in $[\alpha/\mathrm{Fe}]$ at $\mathrm{[Fe/H]}\lesssim -2$ followed by a decline at $\mathrm{[Fe/H]}\gtrsim -2$.

However, a key difference emerges when comparing model-to-model variations. Whereas oxygen shows a strong dependence on the assumed explodability, the other $\alpha$-elements exhibit substantially smaller differences among the models, and their evolutionary tracks remain broadly similar. This can be understood from the progenitor-mass dependence of the yields; the fractional contribution to oxygen production remains relatively high and nearly constant over $M_\mathrm{ZAMS}\sim 20$--$40\,M_\odot$, whereas for many other $\alpha$-elements the contribution decreases with increasing $M_\mathrm{ZAMS}$ in a manner similar to Fe (see Appendix~\ref{sec:comp_yields}). Consequently, even if a larger fraction of high-mass massive stars forms BHs and ceases to contribute to chemical enrichment, the impact on these elements is limited, leading to similar evolutions across models. However, a few elements still show non-negligible model-to-model differences. Magnesium shows somewhat larger model-to-model differences than the other $\alpha$-elements, because its nucleosynthesis production process is similar to oxygen and thus its contribution is relatively flat over $M_\mathrm{ZAMS}\sim 20$--$40\,M_\odot$. However, its sensitivity to explodability remains weaker than in the case of oxygen. In addition, carbon exhibits similar model-to-model differences at low metallicity, since its production in CCSNe is also similar to oxygen and magnesium to some extent. However, the difference diminishes toward solar metallicity; this trend reflects the increasing contribution from AGB stars to carbon enrichment toward higher metallicity. In conclusion, among the elements considered here, oxygen provides the most sensitive diagnostic of explodability changes above $18\,M_\odot$ over the entire metallicity range, and is consequently the most informative abundance to constrain metallicity-dependent explodability with GCE. Conversely, for the other $\alpha$-elements, explodability is less critical, as the different models yield broadly similar evolutionary trends.

Next, we examine individual elements. While most species reproduce the observed trends comparably to oxygen, a few exhibit systematic discrepancies. As discussed above, these elements are largely insensitive to the explodability variations considered here; thus, their discrepancies are not central to our main argument, although we briefly consider possible reasons. For carbon, the discrepancy is minor and may reflect uncertainties in the CCSN yield (see Appendix~\ref{sec:comp_yields}). For titanium, multidimensional effects in the SN explosions are expected to play an important role \citep{maeda2002ApJ...565..405M,maeda2003ApJ...598.1163M}; it has been argued that $[\mathrm{Ti/Fe}]$ should be corrected upward by $\sim +0.45\,\mathrm{dex}$ to take into account such an effect if one-dimensional SN explosion yields are used in GCE study \citep{sneden2016IRONGROUP, kobayashiOrigin2020}. Applying such a correction will substantially improve agreement with observations. For chromium, the observed decrease in $[\mathrm{Cr/Fe}]$ toward low metallicity has been linked to the contribution of hypernovae \citep[see][and Appendix~\ref{sec:comp_yields}]{kobayashiGalactic2006}. Because we do not include hypernova yields in this study, our models are expected to overpredict $[\mathrm{Cr/Fe}]$ relative to the observations.

\begin{figure*}[htb!]
    \includegraphics[width=18cm]{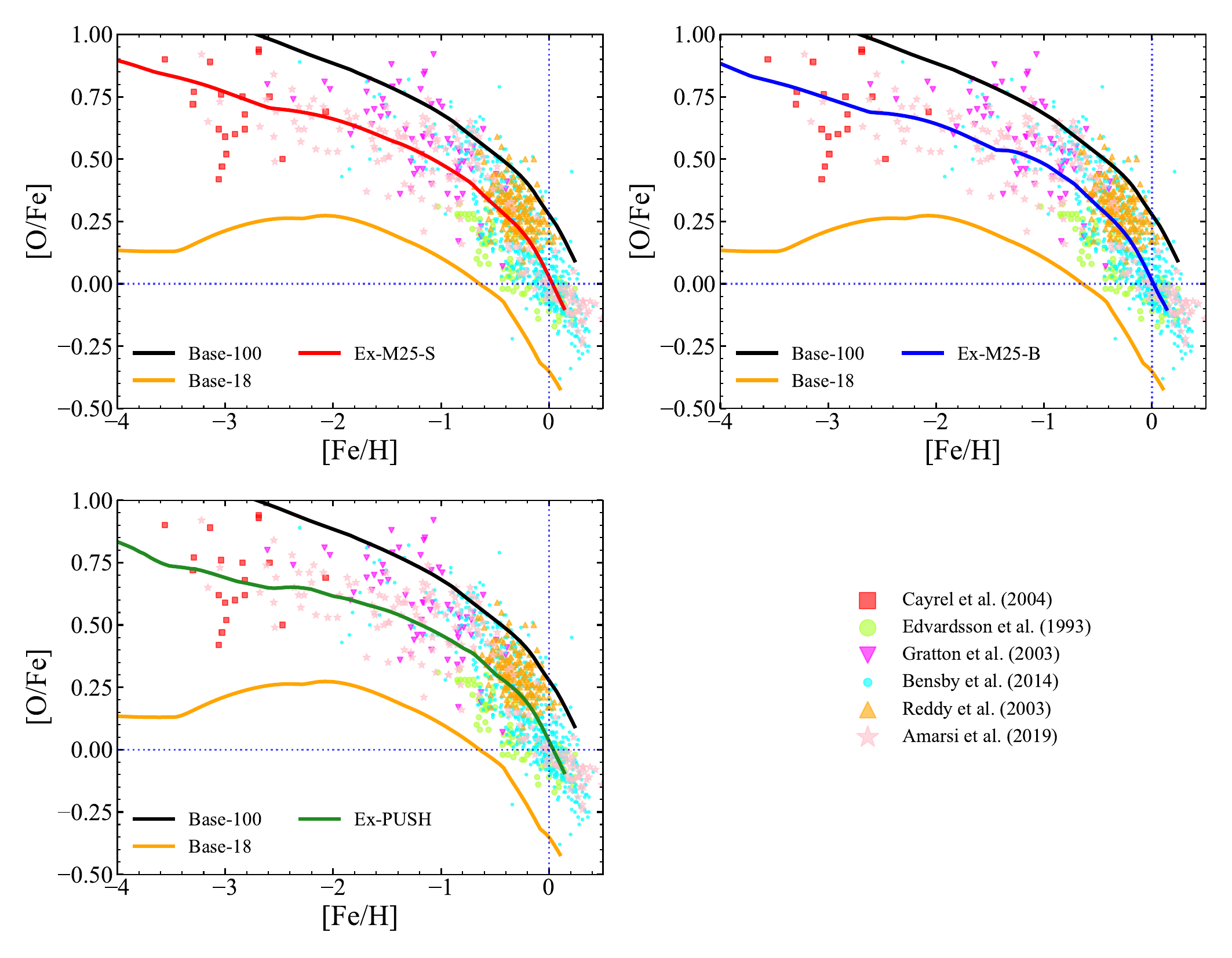}
\caption{
GCE with metallicity-dependent IMF. Same as Figure~\ref{figure:main_O_Fe} but the adapted IMF is metallicity-dependent IMF (Chon IMF) instead of Kroupa IMF.
}
    \label{figure:MD-IMF}
\end{figure*}


\subsection{Influence of Sub-Ch Type Ia SNe}\label{section:sub-CH}
We next evaluate how the inclusion of sub-Ch SNe~Ia modifies the predicted chemical evolution. For sub-Ch SNe~Ia yields, we adopt Table~10 of \citet{leungExplosive2020}, which provides yields for spherical double detonation models with metallicity dependence at $Z=0$, $0.002$ ($\approx0.14Z_\odot$), $0.01$ ($\approx0.7Z_\odot$), and $0.02$ ($\approx1.4Z_\odot$). We assume that sub-Ch and Ch SNe~Ia share the same DTD, and we adopt the same minimum delay time as in our fiducial Ch SNe~Ia model, $50\,\mathrm{Myr}$. When varying the Ch/sub-Ch fraction, we keep all other parameters fixed. The settings of this model are also summarized in Table~\ref{table:GCEparams}.

In Figure~\ref{figure:alpha_elements}, the red dotted lines show the model Ex-M25-S in which the Ch and sub-Ch contributions are assumed to be equal. For the $\alpha$-elements, the predicted evolutions are nearly indistinguishable from the case in which the Ch SNe~Ia contribution is taken to be 100\%, except for titanium, which differs by $\sim 0.1\,\mathrm{dex}$. We therefore conclude that, within the assumptions adopted here, sub-Ch SNe~Ia have a negligible impact on the $\alpha$-element evolution.


\subsection{Yields from BH-forming Events}\label{section:BH_yields}
In the present work, we assume, for simplicity, that BH formation is not accompanied by the ejection of nucleosynthetic products, except for mass loss through stellar winds. A similar zero-yield treatment for non-exploding models was adopted by \citet{jost2025Neutrinodriven} in their GCE calculations based on neutrino-driven CCSN yields. Recent multi-dimensional SN simulations, however, suggest that BH-forming explosions can still eject some metals \citep[e.g.,][]{andersenBlackHoleSupernovae2026, burrowsChannelsStellarmassBlack2025}. However, even in such cases, a large fraction of the CO core is accreted onto the BH for progenitors with $M_\mathrm{ZAMS}\gtrsim 20\,M_\odot$, substantially reducing the yields of metals, including oxygen. For example, in the BH-forming models of \citet{andersenBlackHoleSupernovae2026}, the mass of the ejected Co-core material, i.e., the CO-core mass subtracted by the final remnant mass ($M_\mathrm{CO}-M_\mathrm{rem}$), remains as small as $\simeq0$--$2\,M_\odot$ for progenitors with $20.0< M_\mathrm{ZAMS}/M_\odot\lesssim35.0$ \citep{andersenBlackHoleSupernovae2026,sukhbold2018HighresolutionStudy}. This is substantially smaller than that adopted in the present work for the `successful explosion' case, which give $M_\mathrm{CO}-M_\mathrm{rem}\simeq2$--$8\,M_\odot$ over nearly the same progenitor-mass range \citep{limongi2003EvolutionExplosion,chieffi2004Explosive}. Therefore, for oxygen and most of the other elements discussed in this work, treating BH-forming events as producing no nucleosynthetic ejecta is expected to be a reasonable first-order approximation -- it is not against the idea of having the channel of BH-forming `explosions' seen in recent simulations (see above). Carbon may be an exception, because carbon-rich material is preferentially located in the outer part of the CO core and is therefore more easily ejected than the deeper oxygen-rich material.


\subsection{Implications of Non-Monotonicity in the [Fe/H]--Time Relation}\label{section:multi-infall}
In simple GCE models, especially one-zone models, [Fe/H] is often treated as a monotonic proxy for evolutionary time. This assumption can break down if the gas accretion history is more complex, for example in multi-infall models. In such cases, metal-poor gas accretion can dilute the ISM and make the relation between [Fe/H] and time non-monotonic, so that the same [Fe/H] can correspond to multiple evolutionary stages.

This non-monotonic evolution can produce a loop- or hook-like trajectory in the [O/Fe]--[Fe/H] plane, particularly toward the metal-rich end \citep[e.g.,][]{spitoni2019GalacticArchaeology, spitoni2021APOGEEDR16, hegedus2025ReconstructingMilky}. If star formation continues during this phase, the system can spend additional evolutionary time near high metallicity. As a result, the gas abundance may be more strongly affected by the effective enrichment ratio set by the IMF-averaged CCSN yields and the Fe contribution from SNe Ia around solar metallicity, as discussed in Appendix~\ref{sec:yield_O_Fe}. This effect is expected to be stronger in the simplified models discussed in Section~\ref{sec:simplified_model}, where massive stars that would otherwise supply oxygen at high metallicity fail to explode, thereby shifting the effective enrichment ratio toward lower [O/Fe].


\subsection{Metallicity-dependent Initial Mass Function}\label{section:MD-IMF}

As a test case for a metallicity-dependent IMF, we adopt the Chon IMF proposed by \citet{chon2024Impact}. This IMF, $\psi(M,Z)$, is given by
\begin{align}\nonumber
    \psi(M,Z) &= \psi_0 M^{-\alpha} \left[1 - \exp\!\left(-\left(\dfrac{M}{m_0}\right)^{1.6}\right)\right]\\
    &\times\exp\!\left(-\dfrac{M_l}{M}-\dfrac{M}{M_u}\right),\\
    \alpha &= 2.3 + 0.33\,\log(Z/Z_\odot),\\
    \log m_0 &= 0.2 + 0.45\,\log(Z/Z_\odot),
\end{align}
where $\psi_0$ denote the overall normalization. It becomes more top-heavy at low metallicity and gradually approaches a Salpeter-like IMF \citep{salpeter1955Luminosity} toward solar metallicity.

Because the Kroupa IMF \citep{kroupaVariation2001} used in this study differs substantially from a top-heavy IMF, especially at low metallicity, we expect the overall GCE to change markedly when adopting the Chon IMF. Therefore, reproducing the observational constraints requires retuning the GCE parameters. In this setup, we adopt a star formation timescale of $\tau_\mathrm{s}=2.5\,\mathrm{Gyr}$, an inflow timescale of $\tau_\mathrm{in}=3.5\,\mathrm{Gyr}$, an outflow coefficient of $\epsilon_\mathrm{out}=2.00$, and an initial inflow rate of $\dot{M}_\mathrm{in,0}=71\,M_\odot\,\mathrm{yr}^{-1}$, while keeping the remaining parameters unchanged. The settings of these models are also summarized in Table~\ref{table:GCEparams}.

Figure~\ref{figure:MD-IMF} shows that the models Ex-M25-S, Ex-M25-B, and Ex-M25-PUSH reproduce the observed $[\mathrm{O/Fe}]$--$[\mathrm{Fe/H}]$ trend reasonably well across the full $[\mathrm{Fe/H}]$ range. Compared to Figure~\ref{figure:main_O_Fe}, the enhanced fraction of massive stars at $[\mathrm{Fe/H}]\lesssim-1$ increases $[\mathrm{O/Fe}]$, leading to improved agreement with the observations in the metal-poor regime. While the IMF at low metallicity is still highly uncertain, our result suggests that a top-heavy IMF toward lower metallicity could in general provide a good agreement with the [O/Fe] evolution in GCE, once coupled with the idea of metallicity-dependent explodability. 

Nevertheless, a discrepancy remains: the predicted MDF peak shifts to $[\mathrm{Fe/H}]\sim -0.2$, which is slightly offset from the observed peak. This issue may reflect the limitations of a one-zone treatment and could potentially be alleviated by considering multiple Galactic regions rather than a one-zone model.


\subsection{Missing Supernova Problem}\label{section:missing-Methods}

Finally, we discuss the extent to which metallicity-dependent explodability can modify the predicted cosmic CCSN rate density and thereby contribute to resolving the missing SN problem. If the relation between redshift $z$, used here as a proxy for cosmic time, and the stellar metallicity $Z$ inferred for the Milky Way can be applied to the relation between the cosmic SFR density and the cosmic CCSN rate density, then the predicted cosmic CCSN rate density can be approximated as
\begin{align}
    \dot{n}_\mathrm{cosmic,cc,pred}(z)
    \approx \dot{\rho}_\mathrm{cosmic,*}(z)\,\frac{\dot{N}_\mathrm{cc}(z)}{\dot{M}_*(z)} ,
\end{align}
where $\dot{\rho}_\mathrm{cosmic,*}(z)$ is the cosmic SFR density and $\dot{N}_\mathrm{cc}(z)/\dot{M}_*(z)$ is the CCSN rate per unit SFR implied by the GCE model at the corresponding redshift. For $\dot{\rho}_\mathrm{cosmic,*}(z)$, we adopt the compilation by \citet{madauCosmic2014}.

At $z=0$, $\dot{n}_\mathrm{cosmic,cc,pred}$ is $1.39\times10^{-4}\,\mathrm{SNe}\,\mathrm{yr}^{-1}\,\mathrm{Mpc}^{-3}$ in the model Ex-M25-S. Observationally, the local CCSN rate density has been reported as $6.32^{+0.90}_{-0.81}\times10^{-5}\,\mathrm{SNe}\,\mathrm{yr}^{-1}\,\mathrm{Mpc}^{-3}$ \citep{pessi2025Supernova}, implying a discrepancy by a factor of $\sim 2$. Even if we normalize the peak value in the top panel of Figure~\ref{figure:Metallicity_SNR}, the total CCSN rate in model Ex-M25-S decreases only by $\sim 5\%$ relative to model Base-100, indicating that metallicity-dependent explodability is unlikely to provide a fundamental resolution of the missing SN problem.


\section{Conclusions}\label{sec:conclusion}

In this study, we aimed at placing GCE constraints to the so-called explodability that describes the fate of massive stars -- exploding as successful SNe (to contribute metal enrichment to galaxies like the MW) or collapsing into BHs (without metal enrichment). We incorporated not only its dependence on the initial mass ($M_{\rm ZAMS}$) but also on the metallicity. One motivation lied on the `RSG problem' -- massive RSG progenitors ($M_{\rm ZAMS} \gtrsim 18 M_\odot$) are not found in the local Universe. First, we presented GCE models adopting the recently proposed metallicity-dependent explodability \citep{maltsevExplodability2025,ebinger2019PUSHing,ebingerPUSHing2020}. In addition, we tested a recently proposed metallicity-dependent IMF \citep{chon2024Impact} in which the IMF becomes top-heavy toward low metallicity, as coupled with the metallicty-dependent explodability. From this exercise, we drew the following conclusions:

\begin{enumerate}
\item The GCE model can provide a useful test for explodability of massive stars. Especially useful is the evolution of $[\mathrm{O/Fe}]$ as a function of $[\mathrm{Fe/H}]$; this is relatively insensitive to specific yield sets or SN Ia models adopted in GCE calculations. 
\item These standard metallicity-dependent explodability schemes can provide good agreement with the observed trend. 
\item Within the uncertainty of the explodability prescription, maximizing the formation rate (i.e., the mass range) of BHs leads to disagreement in the $[\mathrm{O/Fe}]$ evolution. 
\item The prescription of the explodability can affect the resulting SN rate. However, once the GCE constraints are placed, the SN rate would not change substantially, therefore it would not remedy the `missing SN problem'. 
\item Introducing a metallicity-dependent IMF can improve the agreement with abundance trends but may still leave residual tension in the MDF.
\end{enumerate}

We thus showed that a important constraint on the explodability can be placed, and the argument can be further generalized -- massive stars up to $M_{\rm ZAMS}\sim 30$--$40\,M_\odot$ must explode at low metallicity as a major contributor to the oxygen enrichment. To capture the characteristic requirements for the explodability from the GCE perspective, as well as further investigate a possible solution to the RSG problem through the explodability, a simple model was constructed in which successful explosions are restricted to $M_{\rm ZAMS}\le 18\,M_\odot$ at $Z\ge Z_\odot$, while explosions up to $M_{\rm ZAMS}=40\,M_\odot$ are allowed below a transition metallicity $Z_{\rm th}$. 

\begin{enumerate}
\item The simplified models required to maintain agreement with the observed oxygen evolution depend sensitively on metal retention. With the standard outflow rate, reproducing the observation values typically requires $Z_{\rm th}\approx Z_\odot$. On the other hand, the constraints can remain satisfied even for $Z_{\rm th}\gtrsim 0.5\,Z_\odot$ without outflow. Although the no-outflow may appear unrealistic, multizone effects may reduce the net metal loss, bringing the solar neighborhood closer to this condition. 
\item The metallicity-dependent explodability can thus potentially solve the RSG problem, without being in tension with GCE constraints. 
\item We demonstrate that the requirement to reproduce the observed oxygen evolution provides a practical route to constraining metallicity-dependent explodability with GCE. 
\end{enumerate}

We note that the explodability of massive stars is still controversial. While higher core compactness has previously been suggested with more difficult explosions \citep{oconnor2011BLACK}, recent work suggests that this trend is not universal and explodability is not that simple \citep{couch2020Simulating,vartanyan2023Neutrino}. Moreover, multi-dimensional explodability results reported in recent years show non-negligible model-to-model differences \citep{maltsevExplodability2025,janka2025longterm}. A more definitive assessment may ultimately require explodability constraints derived from multi-dimensional stellar-evolution and core-collapse calculations.

Several directions for future work are clear. First, the metallicity dependence in the explodability prescriptions adopted here is based on a small number of metallicity points connected by interpolation; it will be important to examine how the results change when explodability is evaluated over a finer grid in metallicity. Second, for simplicity, we characterized explodability primarily by the CO-core mass $M_\mathrm{CO}$ and the metallicity $Z$; in future work, we plan to incorporate the results of more self-consistent multi-dimensional simulations that follow stellar models from the pre-main-sequence phase to their final fate, as such calculations become available. From the observational side, we demonstrated that abundances at super-solar metallicity can help constrain the metallicity-dependence of the explodability; this is promissing since observational constraints in high-metallicity environments, such as the Galactic center, have been rapidly improving \citep[e.g.,][]{nandakumarFirst2025}. 

\begin{acknowledgments}
The authors thank Gaku Kawashima for fruitful discussions and comments. Part of this work was supported by the NAOJ Research Coordination Committee, NINS (NAOJ-RCC-2502-0202), and Fukui Prefectural University. K.M. acknowledges support from JSPS KAKENHI grant (JP24KK0070, JP24H01810, JP24K00682, and JP23H04894). 
\end{acknowledgments}

\appendix

\begin{figure*}[htb!]
    \includegraphics[width=18cm]{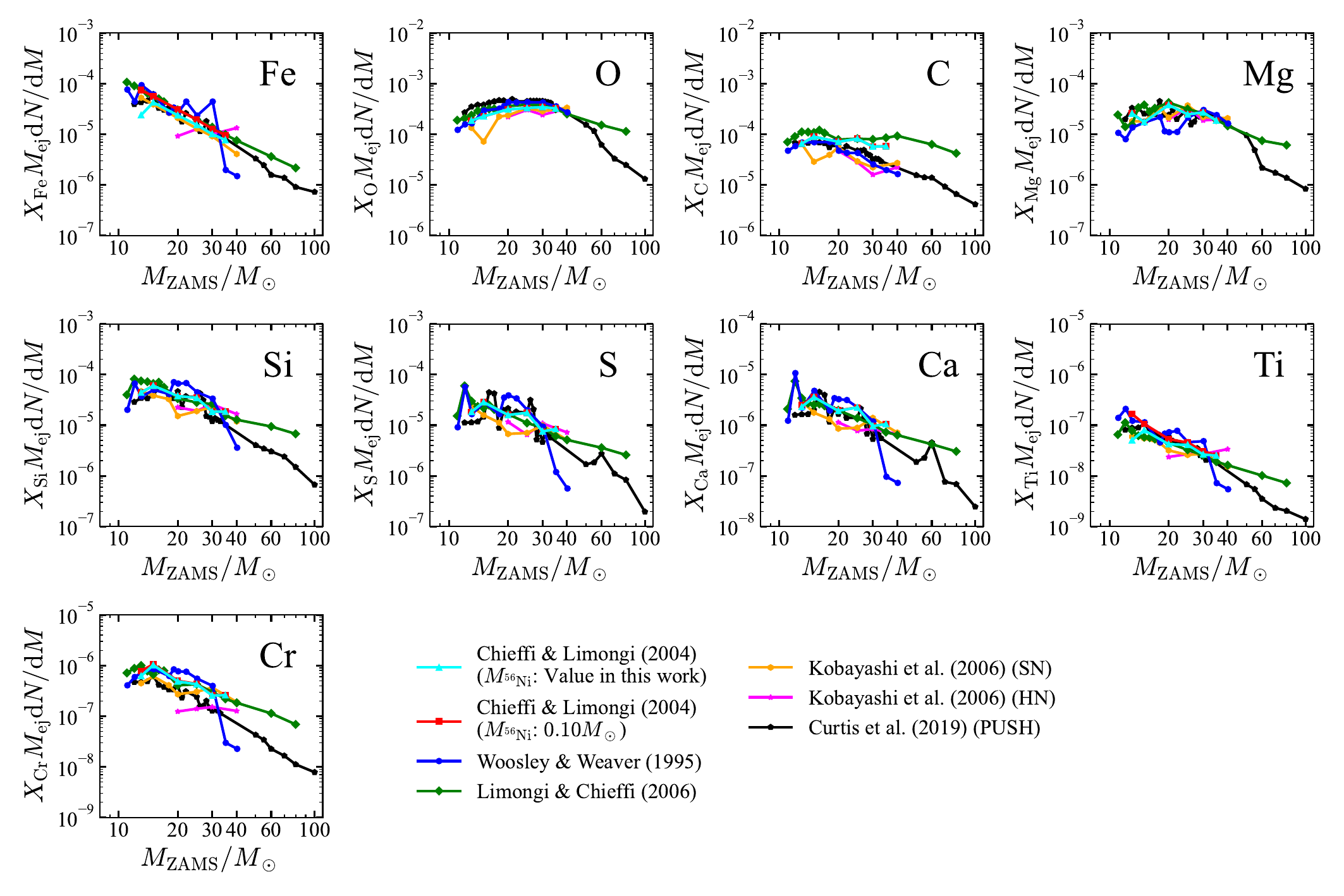}
    \caption{
    Comparison of the elemental yield contributions as a function of ZAMS mass. The horizontal axis denotes the ZAMS mass, and the vertical axis represents the contribution to galactic chemical enrichment per unit ZAMS mass, $X_i\, M_{\mathrm{ej}}\, \mathrm{d}N/\mathrm{d}M$. Each line corresponds to a different published yield set (including both supernova and hypernova models): \citet{chieffi2004Explosive} adjusted using the $^{56}\mathrm{Ni}$ yields in this work (cyan line with upward triangles; in detail see Section~\ref{section:main-method}); \citet{chieffi2004Explosive} assuming $M_{^{56}\mathrm{Ni}} = 0.10\,M_{\odot}$ for all ZAMS masses (red line with squares); \citet{woosley1995Evolution} (blue line with circles); \citet{limongi2006Nucleosynthesis} (green line with diamonds); \citet{kobayashiGalactic2006} supernova (SN) yields (orange line with hexagons) and hypernova (HN) yields (magenta line with stars); and \citet{curtis2019PUSHing} (PUSH; black line with pentagons). All models assume a metallicity of $Z = 0.02$. Data for $M_\mathrm{ZAMS} > 100\,M_\odot$ are omitted in this figure.
    }
    \label{figure:comp_yields}
\end{figure*}


\section{ZAMS-Mass Dependence of CCSN Yields across Different Yield Sets}\label{sec:comp_yields}

Figure~\ref{figure:comp_yields} compares the contribution per $1\,M_\odot$ of each element as a function of ZAMS mass, for several yield sets from the literature. The contribution is defined as $X_i M_\mathrm{ej}\,\dd N/\dd M$, where $X_i$ is the mass fraction of element $i$ in the ejecta and $M_\mathrm{ej}$ is the ejected mass. The factor $\dd N/\dd M$ denotes the number of stars in the interval $[M, M+\dd M]$ when the total stellar mass formed is normalized to $1\,M_\odot$. Writing the IMF as $\psi(M)$, we compute
\begin{align}
    \dv{N}{M}=\dfrac{\psi(M)}{\int_{M_l}^{M_u} M\,\psi(M)\,\mathrm{d}M},
    \label{eq:dNdM}
\end{align}
so that $X_i M_\mathrm{ej}\,\dd N/\dd M$ can be interpreted as the IMF-weighted contribution of element $i$ at each ZAMS mass. The figure includes both supernova yields and hypernova yields.

We first examine the overall trends for each element. As a function of ZAMS mass, two distinct behaviors emerge: (i) elements such as O, C, and Mg, whose contributions remain comparable over $M_\mathrm{ZAMS}\sim 20$--$40\,M_\odot$, and (ii) the remaining elements, whose contributions decrease toward higher masses. Oxygen, in particular, exhibits the strongest tendency toward the flat contribution.

We interpret this dichotomy as a consequence of the mass cut. Heavier species are preferentially synthesized in the deeper layers of the progenitor and are therefore more susceptible to being retained below the mass cut, making their ejecta yields relatively insensitive to increasing $M_\mathrm{ZAMS}$. After IMF weighting, their integrated contribution consequently declines with $M_\mathrm{ZAMS}$, following the decreasing IMF toward higher masses. In contrast, oxygen (and similarly produced $\alpha$-elements) is less affected by the mass cut; its ejecta yield increases with $M_\mathrm{ZAMS}$, which largely compensates for the IMF decline and leads to an approximately flat contribution. This qualitative behavior is observed regardless of metallicities.

Next, we examine the differences in CCSN yields between different yield sets for each element. Because different stellar-evolution and explosion calculations adopt different physical assumptions and calibrations, it is not uncommon for the predicted nucleosynthetic yields to differ substantially between different yield sets. Therefore, when yields and explodability are considered jointly, it is, in principle, preferable to adopt both from a single, internally consistent framework. Nevertheless, some elements are known to be relatively insensitive to the details of the underlying model, and for such elements a practical combination of yields and explodability from different sources can still be adequate.

This study focuses primarily on the GCE of oxygen abundance, and Figure~\ref{figure:comp_yields} indicates that the model-to-model differences in the oxygen contribution are relatively small. The small differences in oxygen production reflect the fact that the oxygen yield is largely determined by the pre-supernova stellar evolution \citep{curtis2019PUSHing,limongi2003EvolutionExplosion} and does not vary substantially among the yield sets. It should also be noted that recent studies have suggested that uncertainties in stellar evolution, such as shell mergers \citep[e.g.,][]{rauscherNucleosynthesisMassiveStars2002}, multi-dimensional effects \citep[e.g.,][]{rizzutiShellMergersLate2024}, and overshooting \citep[e.g.,][]{temaj2024Convectivecore}, may affect oxygen yields. Nevertheless, within the scope of the present comparison, these effects are expected to be secondary compared to the much larger effect of explodability. These discussions suggest that our main conclusions are not strongly dependent on the specific yield set adopted in the present work. The same is true for Fe, which serves as a key reference element in abundance-ratio diagnostics. In contrast, carbon shows a slight difference between supernova yields in different yield sets. Moreover, chromium shows a pronounced difference between supernova and hypernova yields, with the hypernova contribution being substantially smaller than that from supernovae.

\begin{figure*}[htb!]
    \begin{center}
        \includegraphics[width=18cm]{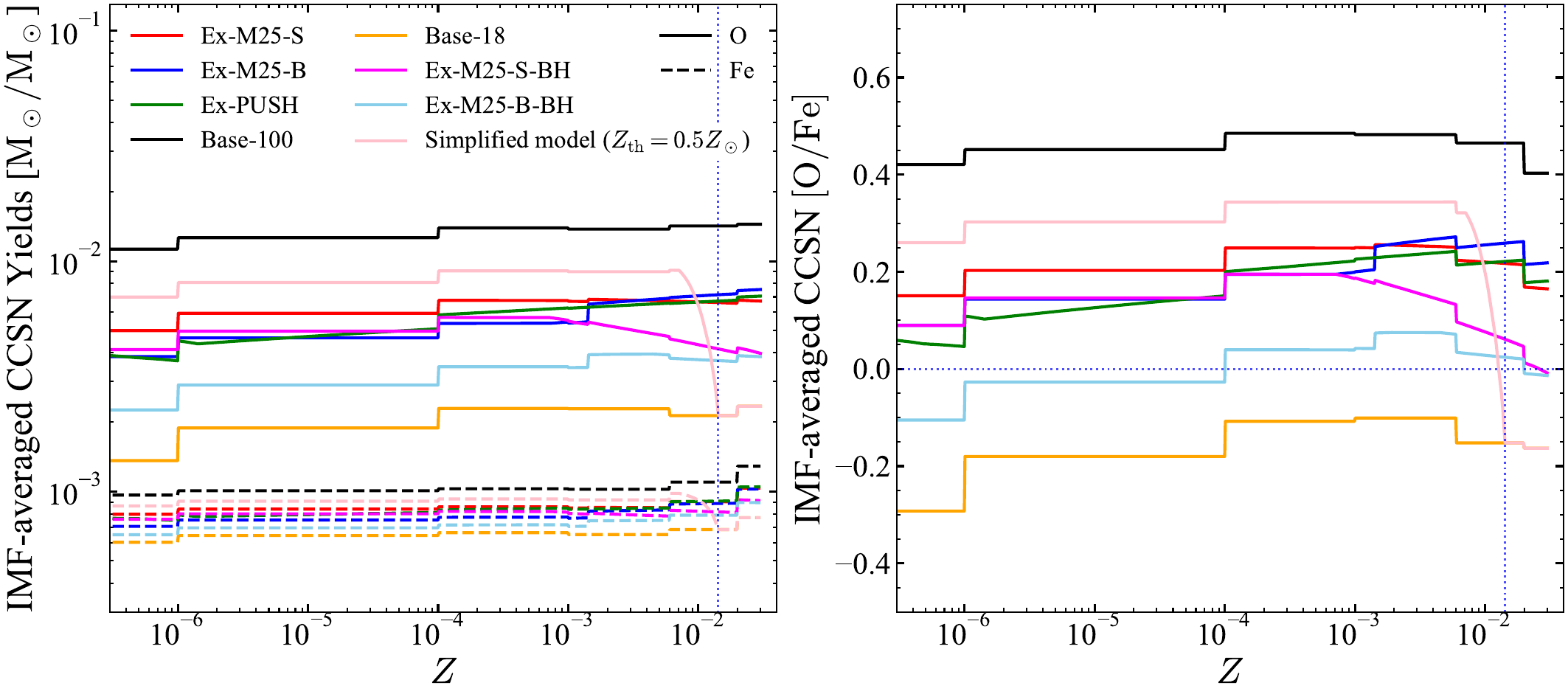}
    \end{center}
    \caption{
    Metallicity-dependent CCSN yields and $[\mathrm{O/Fe}]$ in our models. The left panel shows the IMF-averaged CCSN yields of oxygen (solid) and iron (dashed) obtained when $1\,M_\odot$ of gas is fully converted into stars for each explodability model. The right panel shows $[\mathrm{O/Fe}]$ of the IMF-averaged CCSN yields. The horizontal axis denotes the metallicity in both panels. The colors for the base models and group Ex models are the same as in Figure~\ref{figure:main_O_Fe}, but these of models Ex-M25-S-BH, Ex-M25-B-BH, and the simplified model with $Z_\mathrm{th}=0.5Z_\odot$ are megenta, sky blue, and pink, respectively. The blue dotted vertical lines in both panels denote $Z=Z_\odot$ and the blue dotted horizontal line in the right panel denotes $[\mathrm{O/Fe}]=0$.
    }
    \label{figure:yields_O_Fe}
\end{figure*}


\section{Metallicity-dependent Oxygen and Iron Yields in Different Explodability Models}\label{sec:yield_O_Fe}

To clarify how metallicity-dependent explodability alters the IMF-averaged CCSN yields of oxygen and iron, we compare the IMF-averaged yields of each model as a function of metallicity. The left panel of Figure~\ref{figure:yields_O_Fe} shows that oxygen exhibits substantial model-to-model variations and a clear metallicity dependence induced by explodability, whereas iron varies much less, implying a weaker sensitivity. This contrast reflects the different mass ranges that dominate the IMF-weighted yields (Appendix~\ref{sec:comp_yields}): oxygen receives a large fraction of its contribution from $M_\mathrm{ZAMS}\sim20$--$40\,M_\odot$, while iron does not.

Because long-term enrichment by CCSNe can be approximated by the IMF-averaged yields, the metallicity dependence of the IMF-averaged CCSN yields provides a useful first-order diagnostic of GCE. Accordingly, the right panel of Figure~\ref{figure:yields_O_Fe} captures the relative ordering of $[\mathrm{O/Fe}]$ among models at each metallicity. 

However, it does not fully explain the GCE tracks. At low metallicity, the full GCE calculations generally predict the $[\mathrm{O/Fe}]$ values larger than the IMF-averaged CCSN values, as found in all the models presented in this paper. This is because the system has not yet converged to the IMF average: the longest-lived CCSN progenitors with $8\,M_\odot$ begin to explode only once the metallicity reaches $Z\sim10^{-4.5}$ (roughly $[\mathrm{Fe/H}]\sim-3.5$), so earlier enrichment is dominated by more massive, short-lived stars that typically produce higher $[\mathrm{O/Fe}]$.

At high metallicity, $[\mathrm{O/Fe}]$ predicted by the GCE calculation tends to be lower than the IMF-averaged yields, because of the increasing contribution by SNe~Ia. For Ch~SNe~Ia, the per-event yields are $0.143\,M_\odot$ (O) and $0.749\,M_\odot$ (Fe) \citep{nomotoEvolution1984}, corresponding to $[\mathrm{O/Fe}]=-1.37$; for sub-Ch~SNe~Ia, typical values are $0.060\,M_\odot$ (O) and $0.600\,M_\odot$ (Fe) \citep{leungExplosive2020}, giving $[\mathrm{O/Fe}]=-1.65$. Thus, once SNe~Ia contribute non-negligibly at $[\mathrm{Fe/H}]\gtrsim-2$, $[\mathrm{O/Fe}]$ in the GCE calculation is driven well below the CCSN-only IMF-averaged value. Unless explodability changes extremely sharply or outflows are strongly suppressed (e.g., the Simplified model; Section~\ref{sec:simplified_model}), $[\mathrm{O/Fe}]$ at high metallicity such as solar metallicity is effectively set by a rate-weighted mixture of IMF-averaged CCSN yields (at that metallicity) and SN~Ia yields, with the CCSN-only IMF-averaged $[\mathrm{O/Fe}]$ acting as an approximate upper limit.

\bibliography{Supernovae_Chemical_Evolution}{}
\bibliographystyle{aasjournalv7}

\end{document}